\newif\ifshowchanges
\showchangesfalse 
\ifdefined\hidechanges
    \showchangesfalse
\fi

\newif\ifshowjournal
\showjournalfalse 
\ifdefined\hidejournal
    \showjournalfalse
\fi

\newif\ifshowlinenumbers
\showlinenumbersfalse 
\ifdefined\linenumbersbuild
    \ifnum\linenumbersbuild=1\relax
        \showlinenumberstrue
    \else
        \showlinenumbersfalse
    \fi
\fi

\newif\ifarxiv
\arxivtrue
\ifdefined\arxivbuild
    \ifnum\arxivbuild=1\relax
        \arxivtrue
    \else
        \arxivfalse
    \fi
\fi

\ifarxiv
    \documentclass[11pt,letterpaper]{article}
    \usepackage[margin=1in]{geometry}

\usepackage{multirow}%
\usepackage{dsfont}
\usepackage{mathrsfs}
\usepackage{amssymb,amsmath,amsfonts,amsthm}
\usepackage[parfill]{parskip} 
\usepackage{enumitem} 
\setlist[enumerate,1]{label={(\roman*)}}
\usepackage{url} 
\usepackage{xspace}
\usepackage[most]{tcolorbox}
\usepackage[utf8]{inputenc}
\usepackage[T1]{fontenc}
\usepackage{lmodern}
\usepackage{microtype}
\usepackage{amsmath,amssymb,amsthm,mathtools}
\usepackage{graphicx}
\usepackage{xcolor}
\usepackage[ruled]{algorithm}
\usepackage{algpseudocode}
\usepackage{lineno}
\usepackage{todonotes}
\usepackage[numbers,sort&compress]{natbib}
\usepackage{thmtools}
\usepackage{thm-restate}
\usepackage{placeins}

\newcommand{\deftg}{\ensuremath{\mathcal{G}=(G_1, \ldots, G_L)}}
\newcommand{\tg}{\ensuremath{\mathcal{G}}}
\newcommand{\tq}{\ensuremath{\hat \tg}}
\newcommand{\trp}{{\textsc{TRP}}}
\newcommand{\texp}{\textsc{TEXP}}

\newcommand{\Nat}{\mathbb{N}}

\newcommand{\Wleft}{\ensuremath{W^{\ell}}}
\newcommand{\Wright}{\ensuremath{W^{r}}}
\newcommand{\Wglue}{\ensuremath{W^{+}}}

\usepackage{changepage}
\usepackage{thmtools}
\usepackage{graphicx}

\definecolor{linkcolor}{RGB}{15, 60, 153} 
\usepackage[
colorlinks=true,
linkcolor=linkcolor, 
citecolor=linkcolor,
urlcolor=linkcolor,
]{hyperref}

\usepackage{doi}
\usepackage[nameinlink,capitalize]{cleveref}

\Crefname{figure}{Figure}{Figures}
\Crefname{claim}{Claim}{Claims}
\Crefname{algorithm}{Algorithm}{Algorithms}

\newtheorem{theorem}{Theorem}[section]

\newtheorem{observation}[theorem]{Observation}

\newtheorem{lemma}[theorem]{Lemma}

\theoremstyle{definition}

\theoremstyle{remark}

\newtheorem{claim}[theorem]{Claim}

\newcommand{\cL}{\ensuremath{\mathcal{L}}}

\newcommand{\cP}{\ensuremath{\mathcal{P}}}
\newcommand{\cQ}{\ensuremath{\mathcal{Q}}}

\newcommand{\cT}{\ensuremath{\mathcal{T}}}

\newcommand{\cY}{\ensuremath{\mathcal{Y}}}

\newcommand{\bE}{\ensuremath{\mathbb{E}}}

\allowdisplaybreaks

\providecommand{\email}[1]{\href{mailto:#1}{#1}}

    \newcommand{\onlyconf}[1]{}
    \showjournaltrue
\fi

\ifdefined\journalbuild
    \ifnum\journalbuild=1\relax
        \showjournaltrue
    \else
        \showjournalfalse
    \fi
\fi

\ifshowlinenumbers
    \usepackage{lineno}
\fi

\newcommand\relatedversion{}
\renewcommand\relatedversion{\thanks{The full version of the paper can be accessed at \protect\url{TODO}.}} 

\newcommand{\new}[1]{{\ifshowchanges\color{blue}\fi#1}}
\newcommand{\journal}[1]{{\ifshowjournal\ifshowchanges\color{purple}\fi#1\fi}}

\newcommand{\prooflemequireach}{%
\begin{proof}
    Fix $P_p,P_q\in\cP$, $X\subseteq P_p$, and a time step $t\in[L]$
    with $t+|\cP|\leq L$ and let $\tq$ be the temporal
    quotient graph induced by $\cP$.
    We use nodes and quotient edges for the vertices and edges of $\tq$,
    in order to avoid confusion with the vertices and edges
    of $\tg$. Moreover, we use parts $P \in \cP$
    and the node in $\tq$ representing $P$ interchangeably, e.g.,
    we say $P$ is a node of $\tq$. Since $\tg$ is connected, every snapshot of $\tq$ is connected.
    Hence, by \Cref{lem:reachability}, after $|\cP|$ time steps all
    nodes of $\tq$ are reachable from any fixed start node. In particular, there
    is a temporal walk in $\tq$ from the node $P_p$ to the node $P_q$ starting at
    time step $t$ and arriving by time step $t+|\cP|$.
    Let this temporal walk in $\tq$ be $B_0,B_1,\ldots,B_m$, where $B_0=P_p$ and
    $B_m=P_q$, and let $t\leq t_1<\cdots<t_m\leq t+|\cP|$ be the time
    steps at which the quotient edges $\{B_{j-1},B_j\}$ are used.

    We first consider one time step. Let $Y\subseteq B_{j-1}$. Since
    $\cP$ is equitable in the snapshot $G_{t_j}$, every vertex of
    $B_{j-1}$ has the same number $d_{j-1,j}^{t_j}$ of neighbors in $B_j$, and
    every vertex of $B_j$ has the same number $d_{j,j-1}^{t_j}$ of neighbors
    in $B_{j-1}$. Counting the edges between $B_{j-1}$ and $B_j$ in two ways
    gives
    \[
        |B_{j-1}|d_{j-1,j}^{t_j}
        =
        |B_j|d_{j,j-1}^{t_j}.
    \]
    If $Y'\subseteq B_j$ is the set of vertices reachable from $Y$ by using
    one edge at time step $t_j$, then
    \[
        |Y'|
        \geq
        \left\lceil
            |Y|\cdot \frac{|B_j|}{|B_{j-1}|}
        \right\rceil .
    \]

    Now set $X_0\coloneq X$. For $j=1,\ldots,m$, let $X_j\subseteq B_j$ be the set of
    vertices reachable from $X_{j-1}$ by using an edge at time step $t_j$.
    Applying the inequality above with $Y=X_{j-1}$ gives
    \[
        |X_j|
        \geq
        \left\lceil
            |X_{j-1}|\cdot \frac{|B_j|}{|B_{j-1}|}
        \right\rceil .
    \]
    By induction over $j$, the inequality above implies
    \[
        |X_j|
        \geq
        \left\lceil
            |X|\cdot \frac{|B_j|}{|P_p|}
        \right\rceil
    \]
    for every $j=0,\ldots,m$. Since $B_m=P_q$ and
    $t_m\leq t+|\cP|$, it follows that
    \[
        |R_{t,t+|\cP|}(X)\cap P_q|
        \geq
        |X_m|
        \geq
        \left\lceil
            |X|\cdot \frac{|P_q|}{|P_p|}
        \right\rceil .
    \]
\end{proof}%
}

\newcommand{\prooflemhequireach}{%
\begin{proof}
Choose a constant $c$ large enough for the two bounds proved below, and set
\[
    \Delta_h
    \coloneq 
    \left\lceil
        c\left(\min\left\{h\cdot \frac{n}{|P_i|},hr\right\}+r\right)
    \right\rceil .
\]
Fix a start time $t_1\in[L-\Delta_h]$. 

We first show that $h$ vertices of $P_i$ become reachable within
$O(hn/|P_i|+r)$ time steps starting at $u$
at time step $t_1$. For a part $P_j\in\cP$
and time step $t \in [L]$, say that an
\emph{overflow} occurs in $P_j$ at time step $t_o$ if
\[
    |R_{t_1,t_o}(u)\cap P_j|
    \geq
    \min\left\{
        |P_j|,
        \left\lceil h\cdot \frac{|P_j|}{|P_i|}\right\rceil+1
    \right\}.
\]
As long as not all vertices are reachable, the reachable set grows by at least
one vertex in every time step~(\Cref{lem:reachability}). Hence, by the
pigeonhole principle, an overflow occurs after at most
\[
    \sum_{P_j\in\cP}
    \left(
        \left\lceil h\cdot \frac{|P_j|}{|P_i|}\right\rceil+1
    \right)
    \leq
    \left\lceil h\cdot \frac{n}{|P_i|}\right\rceil+2r
\]
time steps. Let $t_o$ be the first time step at which an overflow occurs, and let
$P_j$ be a part in which it occurs. Set
$T\coloneq R_{t_1,t_o}(u)\cap P_j$. Applying \Cref{lem:equireach} to
$T\subseteq P_j$ and the target part $P_i$ shows that, after another $r$ time
steps,
\[
    |R_{t_o,t_o+r}(Y)\cap P_i|
    \geq
    \left\lceil |T|\cdot \frac{|P_i|}{|P_j|}\right\rceil
    \geq h .
\]
Since every vertex of $Y$ is reachable from $u$ by time step $t_o$, the above inequality implies
that at least $h$ vertices of $P_i$ are reachable from $u$ by time step
$t_o+r$. Thus, the first time step $t'$ at which $h$ vertices of $P_i$ are reachable
satisfies
$t'-t_1 = O(h\cdot \frac{n}{|P_i|}+r)$.

We now show the alternative bound $O(hr)$. Let $X\subseteq P_i$ be the set of
vertices of $P_i$ currently reachable from $u$. Initially, $X=\{u\}$
at time step $t_1$. Suppose
$|X|<h$. Applying \Cref{lem:equireach} to $X$ shows that, after
$r$ time steps, for every part $P_j\in\cP$, at least $\lceil |X|\cdot
\frac{|P_j|}{|P_i|}\rceil$
vertices of $P_j$ are reachable. Since fewer than $h$ vertices of $P_i$ are
reachable, not all vertices of the graph are reachable. Thus, by \Cref{lem:reachability}
after one more time step some new vertex becomes reachable. Let $P_j$ be the
part containing such a new vertex. Then, for some reachable set
$Y\subseteq P_j$, we have 
\[
    |Y| \geq \left\lceil |X|\cdot \frac{|P_j|}{|P_i|}\right\rceil+1
\]
Applying \Cref{lem:equireach} again, now from $P_j$ back to $P_i$, gives after
another $r$ time steps at least
\[
    \left\lceil |Y|\cdot \frac{|P_i|}{|P_j|}\right\rceil
    \geq
    |X|+1
\]
reachable vertices in $P_i$. Hence, within $O(r)$ time steps, the number of
reachable vertices in $P_i$ increases by at least one,
and thus after $O(hr)$ time steps at least $h$ vertices of $P_i$
are reachable.
Thus, the first time step $t'$ at which $O(h)$ vertices of $P_i$
are reachable satisfies $t'-t_1 = O(hr)$.

Taking the minimum of the two bounds concludes the proof.
\end{proof}%
}

\newcommand{\prooflemvertexedge}{%
\begin{proof}
  Since $u_1 \in S_1$ and $u_2 \in S_2$, we know that $ \gamma_1 \coloneqq \chi^2(u_1,u_1) \neq \chi^2(u_2,u_2) \eqqcolon \gamma_2$. 
  Assume towards a contradiction $\chi^2(u_1,v) = \chi^2(u_2,v')$. Then, in particular, 
  \[\left\{\!\!\left\{\left(
      \chi^2(w,v),
      \chi^2(u_1,w)
    \right)
    \mid w \in V
  \right\}\!\!\right\} = 
  \left\{\!\!\left\{
  \left(
      \chi^2(w,v'),
      \chi^2(u_2,w)
    \right)
    \mid w \in V
  \right\}\!\!\right\}
  .\] 
  However, the first multiset contains the entry $(\chi^2(u_1,v),\chi^2(u_1,u_1))$ whose second component is $\gamma_1$, while the second multiset does not have any such entry, a contradiction.
\end{proof}%
}

\newcommand{\prooflemtwlcounts}{%
\begin{proof}
    Since $u$ and $v$ lie in the same fiber, we have
    $\chi^2(u,u)=\chi^2(v,v)$. By the definition of $2$-WL, the multisets
    \[
        \left\{\!\!\left\{
            \left(\chi^2(x,u),\chi^2(u,x)\right)
            \mid x\in V
        \right\}\!\!\right\}
        \quad\text{and}\quad
        \left\{\!\!\left\{
            \left(\chi^2(x,v),\chi^2(v,x)\right)
            \mid x\in V
        \right\}\!\!\right\}
    \]
    are equal.

    Moreover, \cref{lem:vertexedge} yields that if $x \in S'$ and $x' \notin S'$, then $\chi^2(u,x) \neq \chi^2(v,x')$. Hence, the equality of the above multisets must also hold when restricting them to vertices $x\in S'$. Thus, for any fixed color $\gamma\in C$, we obtain
    \[
        |\{x\in S'\mid \chi^2(u,x)=\gamma\}|
        =
        |\{x\in S'\mid \chi^2(v,x)=\gamma\}|
    \]
    and, with a symmetric line of reasoning,
    \[
        |\{x\in S'\mid \chi^2(x,u)=\gamma\}|
        =
        |\{x\in S'\mid \chi^2(x,v)=\gamma\}|.
    \]
\end{proof}%
}

\newcommand{\prooflemtwlwalks}{%
\begin{proof}
    We proceed by induction on $t'-t$. For $t'=t$, the only
    temporal walk starting at time step $t$ and ending at time step $t$ is the
    trivial walk that stays at its start vertex. Hence, the number of such
    walks from $u$ to $v$ is $1$ if $u=v$, and $0$ otherwise. 

    Now let $t'>t$. Every temporal walk from $u$ to $v$ ending at time step
    $t'$ can be decomposed according to the vertex $x$ visited immediately
    before the last step. Thus, the number of such walks is obtained by summing,
    over all $x\in V$, a quantity that depends only on the two colors
    $\chi^2(u,x)$ and $\chi^2(x,v)$: by induction, $\chi^2(u,x)$ determines the
    number of walks from $u$ to $x$ up to the previous time step, and
    $\chi^2(x,v)$ determines whether the final transition to $v$ at time step
    $t'$ is possible.

    Since $\chi^2(u,v)=\chi^2(u',v')$, the definition of $2$-WL gives equality
    of the multisets
    \[
        \left\{\!\!\left\{
            \left(\chi^2(x,v),\chi^2(u,x)\right)
            \mid x\in V
        \right\}\!\!\right\}
        \quad\text{and}\quad
        \left\{\!\!\left\{
            \left(\chi^2(x,v'),\chi^2(u',x)\right)
            \mid x\in V
        \right\}\!\!\right\}.
    \]
    Therefore, the two sums over intermediate vertices are equal. Hence, the
    number of temporal walks from $u$ to $v$ starting at time step $t$ and
    ending at time step $t'$ equals the corresponding number for $u'$ and $v'$.
\end{proof}%
}

\newcommand{\prooflemwalkregularbiregular}{%
\begin{proof}
Let $H\coloneq H_{t,t'}(S,S')$ and
$H'\coloneq \operatorname{supp}(H_{t,t'}(S,S'))$. We first consider the
weighted graph $H$. By \Cref{lem:twlwalks}, all pairs
$\{u,v\}$ with $u \in S$ and $v \in S'$ and the same $2$-WL color $\chi^2(u,v)=\gamma$ have the
same weight $m_{\gamma}$ in $H$.
Fix a vertex $u\in S$. Its weighted degree in $H$ is
\[
    \sum_{\gamma\in C}
    m_\gamma\cdot
    |\{v\in S'\mid \chi^2(u,v)=\gamma\}|.
\]
By \Cref{lem:twlcounts}, the second factor is independent of the choice of
$u\in S$, for every color $\gamma$. Hence, every vertex of $S$ has the same
weighted degree in $H$. Every vertex of $S'$ has the same weighted degree in
$H$ by an analogous argument. Hence, $H$ is biregular.

Finally, an edge of color $\gamma$ is present in the support graph $H'$ if
and only if $m_\gamma>0$. Therefore, the same sums with each positive
$m_\gamma$ replaced by $1$ show that all vertices of $S$ have the same degree
in $H'$ and all vertices of $S'$ have the same degree in $H'$. Hence, $H'$ is
biregular as well.
\end{proof}%
}

\newcommand{\prooflemdoubly}{%
\begin{proof}
By \Cref{lem:walk-regular-biregular}, $H$ is biregular.
Since both vertex classes of $H$ are copies of $S$, the common degree of the
vertices in the right copy of $S$ is also $d$. Therefore, every row and every
column of $P$ contains exactly $d$ entries equal to $1/d$, and hence every row
and every column sums to $1$.
\end{proof}%
}

\newcommand{\prooflemcommontimes}{%
\begin{proof} 
Since the exploration scale is 
always divided by
$f=1+(c-1)/c$ at every recursive call, the first property is immediate. For the second property, note that a
node at depth $i$ is a leaf exactly if $\lceil K_0/f^i\rceil=1$. Thus, the height
$h$ of $\cT$ is $\lceil\log_f K\rceil=\Theta(\log K_0)$ since $f=1+(c-1)/c=2-1/c\geq 3/2$ 
because
$c=C\log |S|\geq 2$.
Additionally, all leaves are exactly at depth $d=h$ by the
first property, and $\cT$ is binary. 

To show the third property, it suffices to show that, for every internal node
$x$ with exploration scale $K_x$, the first time step $t_3\geq t_2$ satisfying
$|R_{t_2,t_3}(v')\cap S|\geq cK_x/f$ is independent of the endpoint $v'$ of
$\Wleft_x$.

\medskip
\noindent \emph{Claim.} If the call represented by a node $x$ starts
at a fixed time step, then it ends at a fixed time step. Moreover, if $x$ is an
internal node, then its right child $x_r$ starts at a fixed time step.

\medskip
\noindent \emph{Proof of the claim.} 
We proceed by induction on the
height of $\cT_x$. If $x$ is a leaf, then it returns the temporal walk
that stays at its start vertex in its prescribed start time step. Thus, its end
time is fixed. Now let $x$ be an internal node, and as
stated in our claim, assume that the call
represented by $x$ starts at a fixed time step. Its left child $x_\ell$ starts
at the same time step. By the induction hypothesis, the walk returned by
$x_\ell$ ends at a fixed time step $t_2$. Let $K'=K_{x_{\ell}}=K_{x_r}=K_x/f$ be
the exploration scale of the two child calls at $x$, and let $v'\in S$ be
the final vertex of the walk returned by $x_\ell$. For any time step $t_3\geq
t_2$, the support graph of the temporal-walk relation from $S$ to $S$ between
time steps $t_2$ and $t_3$ is biregular by \Cref{lem:walk-regular-biregular}.

Hence, $|R_{t_2,t_3}(v')\cap S|$ is independent of $v'$. It follows that the
first time step $t_3\geq t_2$ satisfying
$|R_{t_2,t_3}(v')\cap S|\geq cK_x/f=cK'$ is independent of $v'$ and of all
random choices. Thus, the start time of $x_r$ is fixed. By the induction
hypothesis, the walk returned by $x_r$ ends at a fixed time step. Since the
walk returned by $x$ ends when the walk returned by $x_r$ ends, the end time of
$x$ is fixed as well. This ends the proof of the claim.

\medskip
We can now show our lemma. The root $x_0$ of $\cT$ starts at the fixed time
step $t$. Hence, by the claim, its end time is fixed. Furthermore, whenever an
internal node $x$ starts at a fixed time step, its left child $x_{\ell}$ starts at the same
fixed time step and its right child $x_{r}$ starts at a fixed time step by the claim.
Proceeding from the root $x_0$ towards the leaves therefore shows that every node of
$\cT$ starts at a fixed time step. Then, by the claim, every node also
ends at a fixed time step. 
\end{proof}%
}

\newcommand{\proofclboundf}{%
\begin{proof}
    Since $f=2-1/c$, we have
    \[
        \frac{2}{f}
        =
        \frac{2}{2-1/c}
        =
        \frac{2c}{2c-1}
        =
        1+\frac{1}{2c-1}
        \leq
        1+\frac{1}{c},
    \]
    where the last inequality uses $c\geq 1$. Let $a>0$ be a constant such
    that $d\leq a\log |S|$. By choosing $C\geq a$, we have $d/c\leq 1$.
    Using $1+x\leq e^x$ for $x\geq 0$, we obtain
    \[
        \left(\frac{2}{f}\right)^d
        \leq
        \left(1+\frac{1}{c}\right)^d
        \leq
        e^{d/c}
        =
        O(1).
    \]
\end{proof}%
}

\newcommand{\prooflemnumberofleaves}{%
\begin{proof}
    Recall that at each step of the recursion we divide the current exploration
    scale by $f=1+(c-1)/c$.
    Let $h_x$ be the height of the subtree $\cT_x$. Since $x$ has
    exploration scale $K_x$, a node at depth $d$ in $\cT_x$ has
    exploration scale $K_x/f^d$. Recall that the recursion
    stops once $\lceil K_x/f^d\rceil=1$; it follows that
    $h_x=\lceil\log_f K_x\rceil$.

    By \Cref{lem:common-times}, the subtree $\cT_x$ has
    $2^{h_x}$ leaves. We have $K_x\leq f^{h_x}<fK_x$. Therefore, using
    \Cref{clm:boundf} with $d$ set to $h_x=O(\log |S|)$,
    \[
        2^{h_x}
        =
        \left(\frac{2}{f}\right)^{h_x} f^{h_x}
        =
        O(f^{h_x})
        =
        O(K_x).
    \]
    As $f<2$ and $h_x \geq \log_f K_x$, we have
    $2^{h_x}\geq f^{h_x}\geq K_x$. This gives the matching lower bound, and
    hence $\cT_x$ has $\Theta(K_x)$ leaves.
\end{proof}%
}

\newcommand{\prooflemnumtimesteps}{%
\begin{proof}
Recall that \textsc{PartialExplore} sets $c\coloneq C\log |S|$ for a sufficiently
large constant $C$, sets $f\coloneq 1+(c-1)/c=2-1/c$, and then calls
\textsc{Recurse} with initial exploration scale $K_0\coloneq |S|/c$. Let
$\cT$ be the recursion tree of this call. Recall that a leaf
$y$ of $\cT$ corresponds to a walk $W_y$
that spans one time step and visits one vertex.
By
\Cref{lem:number-of-leaves}, the leaves of $\cT$ contribute
$O(K_0)=O(|S|/\log |S|)$ time steps in total, which is dominated by the
bound below.

We now bound the time steps contributed by all walks $\Wglue_x$ at internal
nodes $x$ of $\cT$. Let $K_d\coloneq K_0/f^d$ be the exploration scale of the
calls at depth $d$. An internal node $x$ at depth $d$ makes two recursive calls
with exploration scale $K_{d+1}=K_d/f$. It first computes the walk $\Wleft_x$,
which ends at some vertex $v'$ at time step $t_2$. The end time $t_3$ of the
connecting part is chosen so that the set of vertices of $S$ reachable from
$v'$ between $t_2$ and $t_3$ contains at least $cK_{d+1}$ possible start
vertices for the walk $\Wright_x$. The walk $\Wglue_x$ is then one temporal
walk from $v'$ to the chosen start vertex of $\Wright_x$. By
\Cref{lem:hequireach}, the walk $\Wglue_x$ spans
$O(qcK_{d+1})=O(qcK_0/f^{d+1})$ time steps.

By \Cref{lem:common-times}, $\cT$ is a binary tree with exactly $2^d$ nodes
at every depth $d$. Hence, all glue walks at depth $d$ together span
\[
    O\left(2^d\frac{qcK_0}{f^{d+1}}\right)
    =
    O\left(\frac{qcK_0}{f}\left(\frac{2}{f}\right)^d\right)
    =
    O(qcK_0)
\]
time steps, where the last equality follows from \Cref{clm:boundf} and
$f>1$. Since $cK_0=|S|$, 
we have $O(qcK_0)=O(q|S|)$ time steps per depth.

Finally, by \Cref{lem:common-times}, $\cT$ has height
$O(\log K_0)=O(\log |S|)$. Hence, the internal nodes contribute
$O(q|S|\log |S|)$ time steps in total.
\end{proof}%
}

\newcommand{\prooflemrecursivetransitionmatrices}{%
\begin{proof}
    Fix any node $x$. We proceed via induction on the height of the subtree
    $\cT_x$. If $x$ is a leaf, then the recursive call represented by
    $x$ returns the temporal walk that stays at its start vertex. Thus, $E_x$
    is the identity matrix. Moreover, the only leaf in $\cT_x$ is $x$
    itself. Thus, $L_{x,x}$ is also the identity matrix. Therefore, both
    matrices are doubly stochastic.

    Now assume $x$ is an internal node of $\cT$,
    and let $x_\ell$ and $x_r$ be its left and right children.
    By the induction hypothesis, $E_{x_\ell}$ and $E_{x_r}$ are doubly
    stochastic, and analogously, all matrices $L_{x_\ell,y}$ with
    $y\in\cT_{x_\ell}$ and all matrices $L_{x_r,y}$ with
    $y\in\cT_{x_r}$ are doubly stochastic. By the third property of
    \Cref{lem:common-times}, the walk returned by $x_\ell$ ends at a fixed time
    step $t_1$, and the walk returned by $x_r$ starts at a fixed time step
    $t_2$, independently of the random choices made by the algorithm. Let
    $P_x\in\mathbb{R}_{\geq 0}^{S\times S}$ be the matrix such that
    $(P_x)_{p,q}$ is the probability that $x_r$ starts at $q$, conditioned on
    $x_\ell$ ending at $p$, for all $p,q\in S$. By the definition of the
    algorithm, $P_x$ is obtained by choosing u.a.r. a vertex of $S$ reachable
    from $p$ between time steps $t_1$ and $t_2$. Therefore, by
    \Cref{lem:doubly}, the matrix $P_x$ is doubly stochastic. The call
    represented by $x$ first executes the call represented by $x_\ell$, then
    performs the random transition represented by $P_x$, and finally executes
    the call represented by $x_r$; its endpoint transition matrix is
    $E_x=E_{x_\ell}P_xE_{x_r}$. As products of doubly stochastic matrices are
    doubly stochastic (\Cref{obs:product-ds}), $E_x$ is doubly stochastic.

    We now consider the leaf-visit matrices $L_{x,y}$ for leaves
    $y\in\cT_x$. We distinguish whether $y$ lies in the left or in the
    right subtree of $x$. First, let $y\in\cT_{x_\ell}$. The call
    represented by $x_\ell$ starts at the same vertex as the call represented
    by $x$. Thus,
    the leaf-visit matrix satisfies $L_{x,y}=L_{x_\ell,y}$,
    which, by the induction hypothesis, is doubly stochastic.
     Now let
     $y\in\cT_{x_r}$. Then the leaf-visit matrix $L_{x, y}$ is obtained
    by first applying the endpoint transition matrix $E_{x_\ell}$ of the left
    child, then the transition matrix $P_x$, and finally the leaf-visit
    matrix $L_{x_r,y}$ inside the right subtree. Thus,
    $L_{x,y}=E_{x_\ell}P_xL_{x_r,y}$. Since all three matrices are doubly
    stochastic (with $L_{x_r, y}$ being doubly stochastic by the induction hypothesis),
    $L_{x,y}$ is doubly stochastic as well.
\end{proof}%
}

\newcommand{\prooflemexpectedleafvisits}{%
\begin{proof} 
Let $x_0$ be the root of $\cT$ and let $y$ be any leaf. For every fixed
vertex $v\in S$, since the start vertex of $x_0$ is chosen u.a.r. from $S$ and
$L_{x_0,y}$ is doubly stochastic by \Cref{lem:recursive-transition-matrices},
we have
\[
    \Pr[y\text{ visits }v]
    =
    \sum_{p\in S}\frac{1}{|S|}(L_{x_0,y})_{p,v}
    =
    \frac{1}{|S|}.
\]
Hence, $\Pr[y\text{ visits a vertex of }D]=|D|/|S|$.

Now, for every leaf $y \in \cY$, let $X_y$ be an indicator variable
with $X_y=1$ exactly if $y$ visits a vertex of $D$. By the probability shown
above, $\bE[X_y]=|D|/|S|$ for every $y\in\cY$. Thus, we can
write $Z_D^{\cY}=\sum_{y\in\cY}X_y$. Applying linearity of
expectation~\footnote{$\bE\bigl[\sum_i
X_i\bigr]=\sum_i\bE[X_i]$}, $\bE[Z_D^{\cY}]
=\sum_{y\in\cY}\bE[X_y] =|\cY|\cdot |D|/|S|$. 
\end{proof}%
}

\newcommand{\prooflemcoincidingleaves}{%
\begin{proof}
    Assume that $\delta$ has been fixed. Recall that $c\coloneq C\log |S|$ and
    $f\coloneq 1+(c-1)/c$, where $C$ is a sufficiently large constant whose
    value will be fixed at the end of the proof. In every non-base-case call
    represented by a node $x$ with exploration scale $K_x$, the two recursive
    child calls have exploration scale $K_x/f$, and the reachable set used to
    start the right recursive call has size at least $cK_x/f$.

    We first bound $\bE[Q_D^x]$ for a fixed internal node $x$
    with a left child $x_{\ell}$ and right child $x_{r}$.
    Let $K'=K_{x_{\ell}}=K_{x_r}=K_x/f$ be the exploration scale passed to each of the
    two children $x_{\ell}$ and $x_{r}$.
    Let
    $m_x\coloneq |\cL_{x_\ell}|=|\cL_{x_r}|$.
    In the following, we condition on a fixed left subtree
    $\cT_{x_\ell}$: the endpoint $v\in S$ of the walk returned by
    $x_\ell$ and the vertex visited by every leaf in $\cL_{x_\ell}$ are
    now fixed.

    Let $P_x$ be the transition matrix from the endpoint of $x_\ell$ to the
    start vertex of $x_r$. By the definition of the algorithm, the start
    vertex of $x_r$ is chosen u.a.r. from a set of at least $cK'$
    vertices. Hence, every entry of the row $(P_x)_v$ is at most
    $1/(cK')$. Now let $z\in\cL_{x_r}$ be any fixed leaf. Since
    the endpoint $v$ is fixed, the distribution of the vertex visited by
    $z$ is given by the row vector $(P_x)_vL_{x_r,z}$. By
    \Cref{lem:recursive-transition-matrices}, the leaf-visit matrix
    $L_{x_r,z}$ is doubly stochastic. Thus, by
    \Cref{obs:product-ds-row}, every entry of
    $(P_x)_vL_{x_r,z}$ is at most $1/(cK')$. Consequently, conditioned on $\cT_{x_{\ell}}$,
    every fixed leaf in the right subtree visits any
    fixed vertex of $S$ with probability at most $1/(cK')$.

    For every $y\in\cL_{x_\ell}$ and
    $z\in\cL_{x_r}$, let $I_{y,z}$ be the indicator variable for
    the event that $\{y,z\}$ is a collision in $D$. If $y$ visits a
    vertex outside $D$, then $I_{y,z}=0$. Otherwise, let $w\in D$ be the
    vertex visited by $y$. 
    We have shown that the probability that $z$ also visits $w$ is at most
    $1/(cK')$, conditioned on $\cT_{x_{\ell}}$.
    Recall that $Z_D^{\cL_{x_\ell}}$
    counts (with multiplicity) the number of leaves in
    $\cL_{x_\ell}$ whose visited vertex lies in $D$.
    Since the right subtree $\cT_{x_r}$ contains $m_x$ leaves, every leaf
    counted by $Z_D^{\cL_{x_\ell}}$ contributes at most $m_x/(cK')$ in
    expectation to $Q_D^x$. Therefore, 
    \[
        \bE\bigl[Q_D^x\mid \cT_{x_{\ell}}\bigr]
        =
        \sum_{y\in\cL_{x_\ell}}
        \sum_{z\in\cL_{x_r}}
        \bE\bigl[I_{y,z}\mid \cT_{x_{\ell}}\bigr]
        \leq
        Z_D^{\cL_{x_\ell}}\cdot \frac{m_x}{cK'}.
    \]

    This inequality holds for every possible outcome of the choices made in
    the left subtree. Averaging over these outcomes and using the law of
    total expectation~\footnote{$\bE[X]=\bE[\bE[X \mid Y]]$} gives
    \[
        \bE[Q_D^x]
        =
        \bE\!\left[
            \bE\bigl[Q_D^x\mid \cT_{x_{\ell}}\bigr]
        \right] 
        \leq
        \frac{m_x}{cK'}\,
        \bE\bigl[Z_D^{\cL_{x_\ell}}\bigr].
    \]
    Applying \Cref{lem:expected-leaf-visits} to
    $\cL_{x_\ell}$ yields
    $\bE[Z_D^{\cL_{x_\ell}}]
    =m_x\cdot |D|/|S|$. Consequently,
    \[
        \bE[Q_D^x]
        \leq
        \frac{m_x^2}{cK'}\cdot\frac{|D|}{|S|}.
    \]

    We now sum the bound over all internal nodes, grouped by their depth in
    $\cT$. By \Cref{lem:common-times}, all nodes at a fixed depth $d$ have the same
    exploration scale $K_d=K_0/f^d$. Moreover, all subtrees rooted at depth $d$
    have the same number $m_d$ of leaves.

    Fix a depth $d$ containing internal nodes. Every node $x$ at depth $d$ has two
    child subtrees, each with $m_{d+1}$ leaves and child exploration scale
    $K_{d+1}$. Since the sets $\cL_x$ for nodes $x$ at depth $d$ partition $\cL$,
    there are $|\cL|/m_d$ nodes at depth $d$. Moreover, $m_d=2m_{d+1}$. Using the
    bound on $\bE[Q_D^x]$ obtained above, the expected number of collisions
    assigned to all nodes at depth $d$ is at most
    \[
        \frac{|\cL|}{m_d}\cdot
        \frac{m_{d+1}^2}{cK_{d+1}}\cdot\frac{|D|}{|S|}
        =
        \frac{|\cL|}{2m_{d+1}}\cdot
        \frac{m_{d+1}^2}{cK_{d+1}}\cdot\frac{|D|}{|S|}
        =
        \frac{|\cL|}{2c}\cdot
        \frac{m_{d+1}}{K_{d+1}}\cdot\frac{|D|}{|S|}.
    \]
    By \Cref{lem:number-of-leaves}, there is a constant $b>0$ such that
    $m_{d+1}\leq bK_{d+1}$ for every depth $d$. Hence, the last expression is at
    most
    \[
        \frac{b|\cL|}{2c}\cdot\frac{|D|}{|S|}.
    \]

    Note that the above bound is independent of the actual depth. By
    \Cref{lem:common-times}, the depth of $\cT$ is $O(\log |S|)$. Let
    $a>0$ be a constant such that every internal node of $\cT$ has
    depth at most $a\log |S|$. Since every collision is assigned to
    exactly one internal node (its lowest-common ancestor), it suffices to
    multiply the above per-depth bound by an upper bound on the depth of the
    recursion tree.
    \[
        \bE[Q_D]
        \leq
        a\log |S|\cdot \frac{b|\cL|}{2c}\cdot\frac{|D|}{|S|}
        =
        \frac{ab}{2C}\,|\cL|\cdot\frac{|D|}{|S|}.
    \]
    Choosing $C$ sufficiently large, we ensure that $ab/(2C)\leq\delta$.
    Therefore, $\bE[Q_D]\leq \delta|\cL|\cdot |D|/|S|$, which proves the lemma.
\end{proof}%
}

\newcommand{\prooflempartialexploration}{%
\begin{proof}
    Let $\cT$ be the recursion tree of the invocation and let
    $\cL$ be its set of leaves. Recall that $Z_D$ counts the leaves
    whose visited vertex lies in $D$, counted with multiplicity, while $Q_D$
    counts the pairs of leaves that visit the same vertex of $D$. Since the
    vertices visited at the leaves are also visited by $W$, it holds for every
    execution that $|V(W)\cap D|\geq Z_D-Q_D$.

    Choose the constant $C$ in the definition of $c=C\log |S|$ sufficiently
    large so that \Cref{lem:coinciding-leaves} holds with, say, $\delta=1/2$.
    Then, by \Cref{lem:expected-leaf-visits,lem:coinciding-leaves},
    $\bE[|V(W)\cap D|] \geq \frac{1}{2}|\cL|\cdot\frac{|D|}{|S|}$. By
    \Cref{lem:number-of-leaves}, the number of leaves satisfies
    $|\cL|=\Theta(K_0)$. Since the outermost invocation uses $K_0=|S|/c$ with
    $c=C\log |S|$, the claim follows for a suitable constant $\gamma>0$.
\end{proof}%
}

\newcommand{\proofclnumchoices}{%
\begin{proof}
    By \Cref{lem:common-times} and \Cref{lem:number-of-leaves}, there are
    $O(|S|)$ many nodes in any recursion tree $\cT$.
    At each node a random choice is made that decides the start vertex
    of the right recursive call. The number of outcomes per choice
    can be upper-bounded by $|S|$.
    Together with starting the initial recursion with a u.a.r. chosen
    vertex of $S$ as the start vertex, we obtain $O(|S|)$
    many choices, each with at most $|S|$ many outcomes.
\end{proof}%
}

\newcommand{\proofclevaluation}{%
\begin{proof}
    Let $\cT$ be a recursion tree. 
    We write $Z_D=\sum_y I_y$, where $I_y=1$ exactly if the leaf $y \in \cL$ visits a vertex
    of $D$. Similarly, we write $Q_D=\sum_{\{y,z\}} I_{y,z}$, where the sum
    ranges over unordered pairs of distinct leaves and $I_{y,z}=1$ exactly if $y$ and
    $z$ visit the same vertex of $D$. By linearity of expectation, it is
    enough to compute the conditional probabilities of these events.

    The probabilities can be obtained by matrix multiplications from the
    transition matrices introduced in the beginning of this section.
    Previously, these matrices have been used only in the analysis of the
    performance of \Cref{alg:sample}, and have not concretely been
    constructed. Now, we require concrete access to these matrices. Clearly,
    each matrix has at most $|S|^2$ entries. Additionally, each matrix can be
    constructed in polynomial time, and we have polynomially many of them. Thus,
    we assume that we have access to all these matrices.

    For a pair of leaves $y,z$, we compute the analogous joint probabilities by
    considering the two root-to-leaf paths together. At each step, we keep track of
    a pair $(u,v)\in S\times S$, where $u$ is the current vertex on the path to
    $y$ and $v$ is the current vertex on the path to $z$. The corresponding
    transition matrices are therefore indexed by pairs of vertices of $S$; they
    have dimension $|S|^2\times |S|^2$ and hence still have polynomial size. The
    common prefix of the two paths is treated only once.

    Thus, $\Phi_D(\sigma)$ can be evaluated by polynomially many matrix
    multiplications. Since the recursion tree has polynomial size and all
    matrices have polynomial dimension, all multiplications take polynomial time.
\end{proof}%
}

\newcommand{\prooflemderandomized}{%
\begin{proof}
    By \Cref{cl:numchoices}, there are only polynomially many random choices
    and each has only polynomially many possible outcomes. By
    \Cref{cl:evaluation}, for each partial fixing $\sigma$ and each possible
    next outcome, the value of $\Phi_D(\sigma)$ can be evaluated in polynomial
    time. We therefore choose, at every step, an outcome maximizing
    $\Phi_D(\sigma)$. This never decreases the conditional expectation, and
    after all choices are fixed the resulting deterministic walk satisfies the
    same lower bound as the randomized construction.
\end{proof}%
}

\newcommand{\proofthmonepartexploration}{%
\begin{proof}
    We proceed in $p$ phases. Let $D_i$ be the set of unvisited vertices before
    phase $i$, with $D_0\coloneq S$ and $i\in\{0,\ldots,p-1\}$. In each phase, we apply
    the derandomized version of \Cref{alg:sample} to obtain a walk that visits at
    least $\gamma |D_i|/\log |S|$ vertices of $D_i$
    (\Cref{lem:partial-exploration}). Thus,
    \[
        |D_{i+1}|
        \leq
        |D_i|\left(1-\frac{\gamma}{\log |S|}\right).
    \]
    Hence, after $p$ phases,
    \[
        |D_p|
        \leq
        |S|\left(1-\frac{\gamma}{\log |S|}\right)^p
        \leq
        |S|\exp(-\gamma p/\log |S|).
    \]
    Choosing $p=\alpha\log^2 |S|$ for a sufficiently large constant $\alpha>0$
    ensures that $|D_p|=O(\log^2 |S|)$. The remaining vertices can be visited
    by repeated applications of \Cref{lem:reachability}. This contributes
    $O(n\log^2 |S|)$ additional time steps.
    As phases are not guaranteed to start at the vertex where the previous phase
    ended, we start each phase $n$ time steps after the previous phase has ended
    and apply \Cref{lem:reachability} to add a temporal walk that connects the two
    phases. Such a walk is guaranteed to exist. In total, all connecting walks span
    $O(n\log^2 |S|)$ time steps.

    Since the walks computed in each phase span $O(q|S|\log |S|)$
    time steps (\Cref{lem:num-time-steps}), we arrive
    at $O(q|S|\log^3|S| + n\log^2 |S|)$ time steps total.
    By \Cref{lem:derandomized}, \Cref{alg:sample} can be derandomized
    in polynomial time. Additional computations, e.g., the application of
    \Cref{lem:reachability}, can easily be done in polynomial time.
\end{proof}%
}

\newcommand{\preliminaries}{
For $k\in\Nat$, we write $[k]\coloneqq\{1,\ldots,k\}$. We denote by
$n$ the number of vertices of the (temporal) graph at hand. 
A temporal graph $\deftg$ with vertex set $V$ and lifetime $L$ is \emph{connected} if every
snapshot $G_t$ for $t \in [L]$ is connected.
A \emph{temporal
walk} in $\tg$ is a
sequence of tuples $(v_0,t_0),\ldots,(v_\ell,t_\ell)$ with $v_i\in V$ and
$t_i\in[L]$ such that $t_0<\cdots<t_\ell$ and, for every $i\in[\ell]$, either
$v_{i-1}=v_i$, or $\{v_{i-1},v_i\}$ is an edge of the snapshot $G_{t_i}$. In the
former case, the temporal walk waits at its current vertex. The \emph{time
span} of the temporal walk is $t_\ell-t_0$. 
For two temporal walks $W$ and $W'$ such that $W$ ends at a vertex $u$ in time
step $t$ and $W'$ starts at $u$ in time step $t$, we denote by $W \circ W'$
the walk resulting from first traversing $W$ and then $W'$.
Throughout the paper, we assume that the lifetime is sufficiently large to accommodate all temporal walks whose existence we assert. If there exists a temporal walk between two vertices $u$ and $v$, we say that $v$ is \emph{reachable} from $u$. We
make use of the following standard result for temporal graphs.

\begin{lemma}[see~{\cite[Lemma~2.1]{ErlebachHK21}}]\label{lem:reachability}
Let $\tg$ be a connected temporal graph with vertex set $V$ and lifetime $L$.
For all \new{$\emptyset \neq X \subsetneq V$}, $t, t' \in [L-1]$
with $t' > t$ \new{and $R_{t,t'}(X) \neq V$}, it holds that
$R_{t,t'}(X) \subsetneq R_{t,t'+1}(X)$.
\end{lemma}

An immediate consequence of the lemma is that every vertex can reach
any other vertex in at most $n$ time steps.

\journal{ 
Let $\deftg$ be a temporal graph with vertex set $V$ and lifetime $L$. Recall
that a temporal graph $\tg$ can equivalently be viewed as its underlying graph
together with an edge coloring encoding the set of time steps an edge is
active. Let $G=(V,E)$ be the underlying edge-colored graph with such an edge
coloring $\lambda\colon E\to C$, and let $\cP=\{P_1,\ldots,P_{\ell}\}$ be a
partition of $V$ into nonempty sets. For $v\in V$, $j\in[\ell]$, and $c\in C$,
let $d_c(v,P_j) \coloneq |\{u\in P_j \mid \{u,v\}\in E \text{ and }
\lambda(\{u,v\})=c\}|$.  We say that $\cP$ is \emph{equitable} if, for all
$i,j\in[\ell]$ and all colors $c\in C$, the value $d_c(v,P_j)$ is the same for
every vertex $v\in P_i$. The \emph{temporal quotient graph} $\tq$ of $\cP$ has
vertex set $\cP$. For two parts $P_i,P_j\in\cP$ and a time step $t\in[L]$, the
edge $\{P_i,P_j\}$ is active in $\tq$ at time $t$ if and only if there exists
an edge $\{u,v\}$ of the underlying graph with $u\in P_i$, $v\in P_j$, and
$t\in\lambda(\{u,v\})$.

Variants of the following two lemmas are known when the vertices of the
temporal graph are partitioned into automorphism orbits; see
\cite[Lemmas~6 and~7]{DogeasEKJMW24}.
The arguments are essentially the same as
in~\cite{DogeasEKJMW24}, we present explicit proofs adapted to our setting of
equitable partitions.

\begin{lemma}[see~{\cite[Lemma~6]{DogeasEKJMW24}}]\label{lem:equireach}
Let $\tg$ be a connected temporal graph with lifetime $L$, and let
$\cP=(P_1,\ldots,P_k)$ be an equitable partition of $\tg$. Then, for
all $P_p,P_q\in\cP$, all $X\subseteq P_p$, and all $t\in[L]$ with
$t+k\leq L$,
\[
    |R_{t,t+k}(X)\cap P_q|
    \geq
    \left\lceil |X|\cdot \frac{|P_q|}{|P_p|}\right\rceil .
\]
\end{lemma}
\prooflemequireach

The following lemma is obtained by a direct application of
\Cref{lem:equireach}. 

\begin{lemma}[see {\cite[Lemma~7]{DogeasEKJMW24}}]\label{lem:hequireach}
Let $\tg$ be a connected temporal graph with lifetime $L$ and vertex set $V$,
and let $\cP=(P_1,\ldots,P_r)$ be an equitable partition of $V$. Let
$P_i\in\cP$, let $u\in P_i$, and let $2\leq h\leq |P_i|$. Then there
exists an integer $\Delta_h = O(\min\{h n/|P_i|,hr\}+r)$
such that, for every $t_1\in[L-\Delta_h]$ it holds that
    $|R_{t_1,t_1+\Delta_h}(u)\cap P_i| \geq h$.
\end{lemma}
\prooflemhequireach

} 
} 

\ifarxiv\title{Temporal Exploration and Rendezvous via Weisfeiler--Leman Colorings}
\else
 \title{Temporal Exploration and Rendezvous via Weisfeiler--Leman Colorings\relatedversion}
\fi
\author{
    {Frank Kammer}\thanks{
        {THM, University of Applied Sciences Mittelhessen
        (\email{frank.kammer@mni.thm.de}, \email{johannes.meintrup@mni.thm.de}). Johannes Meintrup: Funded by the Deutsche Forschungsgemeinschaft (DFG, German Research Foundation) -- 571642628.}
    }
    \and {Sandra Kiefer}\thanks{
        {University of Oxford (\email{sandra.kiefer@cs.ox.ac.uk}).}
    }
    \and {Johannes Meintrup}\footnotemark[1]
}
\date{}
\begin{document}

\ifshowlinenumbers
    \linenumbers
\fi

\maketitle

\ifarxiv\else
\fancyfoot[R]{\scriptsize{Copyright \textcopyright\ 2027 by SIAM\\
Unauthorized reproduction of this article is prohibited}}
\fi

\begin{abstract}
Temporal graphs are graphs whose edge relation changes in discrete time steps.
The Temporal Exploration Problem ($\texp$) asks for a temporal walk that
visits all vertices. We show that every connected temporal graph with $n$
vertices can be explored in $O(qn\log^3 n)$ time steps, where $q$ is the number
of vertex color classes induced by the $2$-dimensional Weisfeiler--Leman
($2$-WL) coloring.
For small $q$, our bound substantially
improves on the known worst-case bound of $\Theta(n^2)$ in general
connected temporal graphs.
A corresponding walk can be
constructed in deterministic polynomial time. The upper bound is complemented
by lower bounds of $\Omega(n \log n)$ when $q$ is constant and $\Omega(qn)$
when $1 \leq q \leq n$.

We also apply our techniques to the Temporal Rendezvous Problem
(\trp{}), in which two agents that do not communicate have to meet on a
vertex of the temporal graph. Our exploration algorithm yields that, with
only polynomial-time computational power, the agents can meet in
$O(n\log^3 n)$ time steps, leaving a gap of only $\log^2 n$ to the known
lower bound.

Dogeas,
Erlebach, Kammer, Meintrup, and Moses Jr.~[ICALP'24] have shown an
$O(rn^{1+\varepsilon})$ bound for $\texp$, for every fixed $\varepsilon>0$,
where $r$ is the number of orbits, and an $O(n^{1+\varepsilon})$ bound for
$\trp$. Both results require superpolynomial computation time. 
Our contribution improves 
upon these results in several ways.
First, the dependence on $n$ in the upper bounds on the number of time steps
is decreased
from $n^{1+\varepsilon}$ to $n\log^3 n$ for both \texp{} and \trp{}.
Second, for $\texp{}$, we bound the number of time steps linearly by the parameter $q$,
which is efficiently computable and upper bounded by $r$.
Third, we provide novel deterministic polynomial-time algorithms for both
\texp{} and \trp{}.
\end{abstract}
\ifarxiv{\newpage}\fi
\section{Introduction}

Temporal graphs are graphs whose edge sets change over discrete time steps. For
surveys and background, see~\cite{CasteigtsFQS12,Michail16}. Formally, a
\emph{temporal graph} $\deftg$
with \emph{lifetime} $L$ is a sequence of graphs, called \emph{snapshots}, on a
common vertex set $V$. 
The graph obtained by the union over all
snapshots is called the \emph{underlying graph}. 
In each time step, a \emph{temporal walk} either stays at the current vertex or traverses an edge present in the current snapshot.
One of the most prominent research directions regarding temporal graphs is the
study of problems involving temporal walks. Examples include temporal
reachability~\cite{AkridaMNRSZ20,CasteigtsCS24,Doring25,ErlebachOM25,WenHZQZL20,WuCYSYW16},
temporal connected components~\cite{BalevSS24,BeckerCCKRRZ26,DeligkasDEGST25}, temporal
spanners~\cite{CasteigtsPS21,CasteigtsRRZ24,KuritaMSU25}, and \emph{Temporal Exploration (\texp{})},
which was first introduced by Michail and Spirakis~\cite{MichailS16}.

The decision variant of \texp{} asks whether a temporal walk exists that visits
all vertices. Deciding $\texp$ is NP-hard, and the corresponding optimization
problem is hard to approximate~\cite{MichailS16,ErlebachHK21}. Even on heavily
restricted graph classes, the problem remains hard~\cite{AkridaMSR21}. For
\emph{connected temporal graphs}, that is, when every snapshot
is connected, Erlebach,
Hoffmann, and Kammer~\cite{ErlebachHK21} showed that $\texp$ can be
solved by a walk spanning $O(n^2)$
time steps and that this bound is
tight.\footnote{Here and in the following,
we let $n \coloneq |V|$.} 

The lower-bound results have motivated extensive work on identifying
restricted settings in which the quadratic barrier can be broken.
Subquadratic upper bounds are known for connected temporal graphs with bounded-degree
snapshots~\cite{ErlebachKLSS19,ErlebachS18,BastideGMR25}, 
planar underlying graphs and underlying graphs of bounded pathwidth or
treewidth,
and cycles with
few chords~\cite{AdamsonGMZ22,BodlaenderV19,ErlebachHK21}. Further
bounds
are known for highly structured graph classes, such as cacti and
$(2\times k)$-grids~\cite{Adamson25,AdamsonGMZ22,ErlebachS22b,IlcinkasKW14},
and for variants such as $k$-edge-deficient temporal graphs, temporal graphs
with regularly or probabilistically present edges, and word-representable
temporal graphs~\cite{AkridaMSR21,ErlebachKLSS19,ErlebachS22a}.
A recent result by
Bastide, Groenland, Michel, and Rambaud~\cite{BastideGMR25} gives a unifying
bound of $O(n^{3/2}\sqrt{D\log n})$ in terms of the average temporal maximum
degree $D$. In particular, their result yields $O(n^{3/2}\sqrt{d\log n})$ when every
snapshot has maximum degree at most $d$, and $O(n^{3/2}\sqrt{k\log n})$ when
the underlying graph has average degree at most $k$; the latter also improves
the known bounds for planar underlying graphs and for underlying graphs of
bounded treewidth. The problem has also been studied in probabilistic
settings~\cite{BaguleyGKSSZ25}, where each snapshot is a random spanning tree.
In this model, temporal graphs can be explored in $O(n^{3/2})$ time steps with
high probability.

Despite extensive progress, temporal exploration in connected temporal
graphs is still far from well understood. While many natural restrictions are
known to admit subquadratic exploration, the best upper and lower bounds often
remain widely separated. For instance, for bounded-degree snapshots,
bounded-degree underlying graphs, planar underlying graphs, and snapshots being
cycles, the best lower bound is still only $\Omega(n\log
n)$~\cite{ErlebachHK21}, whereas the best upper bounds remain around $n^{3/2}$
(up to logarithmic factors)~\cite{BastideGMR25}. Even
highly structured cases are not completely settled: for temporal graphs whose
underlying graph is a $(2\times k)$-grid, the known bounds still differ by a
polylogarithmic factor~\cite{ErlebachHK21}. Thus, while the quadratic barrier
can be broken in many settings, the fine-grained behavior of temporal
exploration remains open for several fundamental graph classes.

Recently, Dogeas, Erlebach, Kammer, Meintrup, and
Moses Jr.~\cite{DogeasEKJMW24} showed that temporal exploration in connected
temporal graphs can be parameterized by the number~$r$ of automorphism orbits, that is, by the equivalence classes of vertices under
automorphisms of the graph. For every fixed $\varepsilon>0$, they obtain a
temporal walk that visits all vertices and spans $O(rn^{1+\varepsilon})$ time
steps, together with lower bounds of \new{$\Omega(n\log n)$} when $r$ is constant and
$\Omega(rn)$ for $1 \leq r \leq n$. The known deterministic construction of this
bound relies on exhaustive search over automorphisms, whose number can be as
large as $n!$. The extended version~\cite{DogeasEKJMW24arxiv} gives a randomized polynomial-time algorithm,
but with a weaker bound of $O((n^{5/3}+rn)\log n)$ time steps for exploring a single orbit.

The same work introduces the Temporal Rendezvous Problem~($\trp$),
a temporal variant of Alpern's rendezvous
problem~\cite{Alpern76,Alpern95,Alpern02,AlpernG06}, which has been studied
extensively in static graphs; see the survey~\cite{Pelc12}.
\new{For more recent results on gathering (i.e., multi-agent
rendezvous), see~\cite{BouchardDL18,BouchardDP20,DevismesDL26}.
Related work also studies treasure hunt~\cite{BouchardDLP21,TaShmaZ14}
and other rendezvous variants~\cite{BhagatP22,BouchardBDDP17,BouchardDPP20,BourreauNN26,GanczorzJPS25}.}
In \trp{},
two agents have full knowledge of the graph but do not share a common
labeling of the vertices. After being placed at starting vertices by an
adversary, they have to meet at some vertex (at a synchronized time step).
As
shown in~\cite{DogeasEKJMW24}, the agents can always meet after
$O(n^{1+\varepsilon})$ time steps in connected temporal graphs, for every fixed
$\varepsilon>0$, and a lower bound of $\Omega(n\log n)$ holds.

We study both temporal exploration and temporal
rendezvous through a weaker and efficiently computable notion of symmetry. Our equivalence classes are the $2$-WL fibers, defined by
vertices being of the same color after an execution of the $2$-dimensional
Weisfeiler--Leman algorithm~\cite{WeisfeilerL68}. 

The \emph{$k$-dimensional Weisfeiler--Leman algorithm} ($k$-WL)~\cite{ImmermanL90} is a
combinatorial algorithm that iteratively refines a coloring of the $k$-tuples
of vertices of the input graph $G=(V,E)$. The $1$-dimensional version is also widely known as
\emph{Color Refinement}. Starting with an initial vertex coloring $\lambda$,
for example $\lambda(v) \coloneqq 1$ for all $v \in V$, and setting
$\chi^1_0\coloneq \lambda$, the coloring is updated, for every $i\geq 1$ and every $v\in
V$, by
\[
\chi^1_i(v)
\coloneq 
\left(
  \chi^1_{i-1}(v),
  \left\{\!\!\left\{
    \chi^1_{i-1}(u) \mid \{u,v\} \in E
  \right\}\!\!\right\}
\right),
\]
where the second component of the tuple denotes a multiset of colors. Hence, the color of a vertex in round $i$ consists of its own color in round $i-1$ and the multiset of colors of its neighbors in round $i-1$. Thus, the vertex partition induced by $\chi^1_i$ is at least as fine as
the one induced by $\chi^1_{i-1}$. The process \emph{stabilizes} at some
iteration $i$, where we have $\chi^1_{i}(u) = \chi^1_{i}(v)$ if and only if
$\chi^1_{i+1}(u) = \chi^1_{i+1}(v)$ for all vertices $u$ and $v$. Letting $i^*$ be
the value of $i$ where this first happens, we set $\chi^1 \coloneqq
\chi^1_{i^*}$.

For dimension $k \geq 2$, the algorithm proceeds analogously on
ordered vertex $k$-tuples. We only define the procedure for $k=2$, as it is the
only variant we require in this work. The initial coloring
$\chi^2_0$ of $V^2$ encodes the adjacency and equality type of each
pair of
vertices. For example, $\chi_0^2(u, u) = 0$, $\chi_0^2(u, v)=-1$ if
$\{u, v\}$ is not an edge,
and $\chi^2_0(u, v)=1$ if $\{u, v\}$ is an edge.
Then, for every $i\geq 1$ and every $(u,v)\in V^2$,
\[
\chi^2_i(u,v)
\coloneq 
\left(
  \chi^2_{i-1}(u,v),
  \left\{\!\!\left\{
    \left(
      \chi^2_{i-1}(w,v),
      \chi^2_{i-1}(u,w)
    \right)
    \mid w \in V
  \right\}\!\!\right\}
\right).
\]
Analogously to the case of $k=1$, the algorithm terminates at the first
iteration $i^*$ when the partition induced by the coloring stabilizes. Again,
the coloring $\chi^2 \coloneqq \chi^2_{i^*}$ induces a partition of the
vertices into the \emph{$2$-WL fibers} (or simply the \emph{fibers}) of the
graph: two vertices $u,v \in V$ belong to the same fiber if and only if
$\chi^2(u,u) = \chi^2(v,v)$. While the above procedures are defined for static
graphs, the algorithm can be generalized readily to work on temporal graphs by
taking as the input graph the underlying graph $G$ of the temporal graph
$\deftg$ and including in the initial coloring the labeling of each edge by the
set of time steps in which it is present.
\new{Given such an edge coloring (or any edge coloring for that matter), the $2$-WL fibers
can be computed in time $O(n^3\log n)$~\cite{ImmermanL90,ImmermanS19}.}

For $k,k'$ with $1\leq k' < k$, it follows straight from the definition of the algorithm that $k$-WL captures all the information that $k'$-WL captures, and hence, the partition into the $k$-WL fibers
refines the partition into the $k'$-WL fibers. Moreover, the
partition into automorphism orbits always refines the partition into the $k$-WL
fibers since $k$-WL colors are invariant under automorphisms. Thus, WL fibers
provide polynomial-time computable relaxations of automorphism orbits.
This approximation of symmetry makes WL-type refinement a central tool in
practical graph-isomorphism algorithms
\cite{AndersS21,DargaLSM04,JuntillaK07,JuntillaK11,mckay81,McKayP14} and in
theoretical approaches to the graph-isomorphism problem \cite{ArvindKRV17,Babai16,BabaiES80,CaiFI92,ImmermanL90,KieferSS21}. The dimension of the algorithm required to capture the isomorphism type of a graph has by now
become a common measure for the structural complexity of the
graph~\cite{Grohe17,Kiefer20} and also of other objects \cite{0001GR24}. 
Most notably, the WL algorithm is a core ingredient in Babai's celebrated
quasipolynomial-time isomorphism and automorphism algorithms~\cite{Babai16,Babai19}.
Beyond pure isomorphism testing, the algorithm has surprising connections to and applications in various areas, such as mathematical logic~\cite{CaiFI92,KieferSS21}, proof complexity~\cite{AtseriasM13}, homomorphism counting \cite{DellGR18,Dvorak10}, graph connectivity~\cite{KieferN22}, satisfiability checking~\cite{DargaLSM04}, machine learning~\cite{MorrisRFHLRG19,ShervashidzeSLMB11,XuHLJ19}, and educational support systems~\cite{SchmellenkampZA25}, all
of which contribute to making it a rich and active topic of research. To the
best of our knowledge, our work presents the first use of the WL algorithm to
obtain complexity bounds for reachability problems in temporal graphs.

\subsection{Our Contribution}

Our work studies the behavior of $\texp$ and $\trp$ when
parameterized by the number $q$ of $2$-WL fibers. Common WL-based measures are the
dimension required or sufficient for the algorithm to distinguish
non-isomorphic graphs 
\cite{Grohe12,GroheK19,GroheN23,GuoGP24,KieferN22,KieferN24,KieferPS19,LichterRS26},
the number of iterations \cite{GroheK21,GroheV06,KieferM26,KrebsV15,BergeremGKO23}, or the maximum size of a color class computed by the
algorithm \cite{FuhlbruckKV21}. 
In contrast, our
parameter $q$ directly measures the coarseness of the resulting vertex partition. 
\new{
Static graphs with small $q$
include vertex-transitive graphs (such as Cayley graphs, circulant graphs,
and hypercubes) and distance-regular graphs, all of which have $q=1$. In particular, strongly regular graphs have one fiber
despite potentially having unboundedly many automorphism orbits.
Appropriate temporal variants of these graphs retain $q=1$.
}

\begin{restatable}{theorem}{thmExploration}
    \label{thm:exploration}
    Let $\deftg$ be a connected temporal graph with vertex set $V$ and
    sufficiently large lifetime. Let $\cQ$ be the partition of $V$ into
    its $2$-WL fibers, and let $q \coloneqq |\cQ|$. 
    There exists a temporal walk $W$ that visits all
    vertices of $V$ and spans $O(qn\log^3 n)$ time steps.
    Moreover, $W$ can be computed deterministically in polynomial
    time.
\end{restatable}

Recall that $2$-WL is defined on pairs of vertices. The final coloring encodes information about the number of (specific edge-colored) walks between every pair of vertices. Thus, when considering temporal graphs as edge-labeled static graphs with the temporal edge relations encoded in the initial edge coloring, the final $2$-WL color of a vertex pair captures the counts of specific temporal walks between them. This makes the algorithm a suitable tool for problems relating to
temporal walks, and \Cref{thm:exploration} identifies $2$-WL as a natural
structural tool for temporal exploration.

At the same time, $2$-WL equivalence of vertices is a much weaker symmetry notion
than automorphism orbits. In fact, our main technical contribution is to show that the weak
symmetries supplied
by $2$-WL suffice to obtain tight bounds for
$\texp$ up to logarithmic factors. 
To this end, we construct randomized temporal walks based
on the partition into $2$-WL fibers and analyze the distribution of these walks.
Intuitively, the structure imposed by $2$-WL allows us to maintain balanced
distributions
during the randomized construction, ensuring that, in expectation, no vertex is visited
substantially more often than the others.
This enables us to show that, in expectation, the randomized walks that we construct visit
many distinct vertices within a sufficiently short time span. 
Interestingly, we can 
derandomize the 
construction yielding a polynomial-time deterministic algorithm.

We first apply the described approach to explore a fixed fiber $S$ (\Cref{thm:one-part-exploration}).
Afterwards, exploring all fibers one by one yields \Cref{thm:exploration}.

\begin{restatable}{theorem}{thmOnePartExploration}
    \label{thm:one-part-exploration}
    Let $\deftg$ be a connected temporal graph with vertex set $V$ and
    sufficiently large lifetime. Let $\cQ$ be the partition of $V$ into
    $2$-WL fibers, and set $q \coloneqq |\cQ|$. For every
    $S\in\cQ$, there exists a temporal walk $W$ that visits all
    vertices of $S$ and spans $O(q|S|\log^3 |S|+n\log^2 n)$ time steps. Moreover,
    $W$ can be computed deterministically in polynomial time.
\end{restatable}

\Cref{thm:exploration} improves the related result in~\cite{DogeasEKJMW24} in
several ways. Their result gives a bound of $O(rn^{1+\varepsilon})$ time steps,
for every fixed $\varepsilon>0$, where $r$ is the number of automorphism
orbits. Recall that $2$-WL colors are isomorphism-invariant,
hence $q\leq r$. Our bound is in
terms of a parameter $q\leq r$, improves the dependence on $n$ from
$n^{1+\varepsilon}$ to $n\log^3 n$, and comes with a deterministic
polynomial-time construction. 
On a technical level the strategy of~\cite{DogeasEKJMW24}
highly exploits the perfect symmetry information provided by the automorphism group of the graph.
The tools provided by the automorphism group cannot be adapted to $2$-WL fibers,
and thus we require a novel approach. We discuss the intricacies
of our approach in \Cref{sec:proof-sketch}.

    The following lower bounds are implied by known
constructions from~\cite{DogeasEKJMW24,ErlebachHK21}.

\begin{restatable}{theorem}{corLowerbound}
    \label{cor:lowerbound}
    There exist connected temporal graphs $\deftg$ with vertex set $V$ such that,
    letting $\cQ$ denote the vertex partition into $2$-WL fibers and $q
    \coloneqq |\cQ|$, \texp{}
    requires $\Omega(n \log n)$ time steps whenever $q$ is constant. 
    For all $q$ with $1\leq q \leq n$,
    there is a connected temporal graph
    on which \texp{}
    requires $\Omega(qn)$ time steps.
\end{restatable}

For a self-contained presentation of the next result, we now
formally introduce 
    the \emph{Temporal Rendezvous Problem} (\trp{}).
Let $\deftg$ be a temporal graph with vertex set $V$
and lifetime $L$. Two agents $a_1$ and $a_2$ are placed at respective start
vertices $u_1,u_2\in V$ in time step $1$.
The agents run possibly different programs $p_1$ and $p_2$, respectively.
Using these programs, they compute and follow respective temporal walks
$W_1$ and $W_2$. The goal is that the agents meet at the same vertex in the
same time step at least once during these walks. The agents have full
information about the structure of $\tg$, but they cannot communicate and do not see each
other's location. Moreover, they do not have access to a common vertex
labeling: equivalently, each agent sees an isomorphic copy of $\tg$, say
$\tg_1$ and $\tg_2$, with possibly unrelated vertex labels. The start vertices
and the two labelings are chosen by an omnipotent adversary after the programs
$p_1$ and $p_2$ have been fixed. Without consistent vertex labels, the agents
cannot use trivial strategies such as agreeing to meet at the lowest-labeled
vertex. Since the temporal graphs we consider are connected, assuming
sufficiently large lifetime, there always exist temporal walks that make the
agents meet. A solution to $\trp$ is a pair of programs $(p_1,p_2)$ for the two
agents. We are interested in the number of time steps required until the agents
meet and in the complexity of the computation of the walks.

By an application of \Cref{thm:one-part-exploration}, we obtain the following.

\begin{restatable}{theorem}{thmRendezvous}\label{thm:rendezvous} 
    In the \trp{} setting,
    on connected temporal graphs with sufficiently large lifetime,
    the two agents can always meet within $O(n\log^3 n)$ time steps using polynomial computation time.
\end{restatable}

\new{
Note that here, in contrast to \Cref{thm:exploration},
the parameter $q$ has no impact on the time steps.}
Our result for $\trp$ leaves only a factor of $O(\log^2 n)$ to the lower bound of $\Omega(n\log n)$~\cite{DogeasEKJMW24}.
It also improves on the previous upper bound of $O(n^{1+\varepsilon})$ time
steps~\cite{DogeasEKJMW24}. Moreover, our strategy can be implemented by
agents with polynomial-time computational power whereas the previous
approach relies on automorphism orbits and a brute-force enumeration
of temporal walks.%
\onlyconf{\new{\ 
Next, we present an overview of the techniques that we use to obtain our main results.
    Following that, 
   in \Cref{sec:conclusion}, we present a conclusion together 
   with an outlook.
}}
\journal{\ 
The remainder of this work is structured as follows. In
\Cref{sec:proof-sketch}, we sketch the main ideas and techniques to prove our
results and highlight the intricacies that come with using the structural
information provided by $2$-WL rather than the full symmetry information from
automorphism orbits.In \Cref{sec:preliminaries}, we give the remaining
preliminariesand prove basic reachability facts used throughout. In
\Cref{sec:twowl}, we collect the properties of $2$-WL that we use for temporal
walks. Next, in \Cref{sec:algorithm}, we prove \Cref{thm:exploration,thm:one-part-exploration,thm:rendezvous}.
\Cref{sec:lower} describes the lower-bound constructions underlying
\Cref{cor:lowerbound}. We conclude our work with an outlook in \Cref{sec:conclusion}.}

\section{Detailed Technical Overview}\label{sec:proof-sketch}
\onlyconf{\new{We begin by introducing some preliminaries.%
\preliminaries}}%
For the remainder of this section, let $\deftg$ be a connected temporal graph
with vertex set $V$ and lifetime $L$. We introduce some additional notation.
For a set $U\subseteq V$ and time steps $t,t'\in\{1, \ldots, L\}$ with $t'\geq
t$, we write $R_{t,t'}(U)$ for the set of vertices reachable by a temporal walk
that starts at some vertex of $U$ at time step $t$
and ends no later than time
step $t'$. If $U$ is a singleton set $\{u\}$, we set $R_{t,t'}(u) \coloneqq
R_{t,t'}(\{u\})$.

A temporal walk that visits all vertices is usually constructed incrementally.
At the beginning, when almost all vertices are still unvisited, making progress
(i.e.,\ making sure to visit previously undiscovered vertices) is easy. Even a
simple greedy strategy, such as moving to the earliest reachable unvisited
vertex, can visit many new vertices quickly. 
Over time, however, such a greedy
strategy becomes less efficient. Once many vertices have already been visited,
the walk may have to spend a lot of time steps before it reaches a fresh vertex. 

It is well-known that one can visit any set $D \subseteq V$ of vertices
in $|D|\cdot n$ time steps (\Cref{lem:reachability}),
and that in general temporal graphs, this is the best one can hope
for~\cite{ErlebachHK21}.
However, in certain temporal-graph classes,
one can break this barrier. A common method is to
first prove an intermediate progress
guarantee. 
The approach is to show that, for any subset $D\subseteq V$ of the vertices,
there is a temporal walk that visits a guaranteed fraction of
$D$, in much less than $|D|\cdot n$ time steps.
Such
a statement is often easier to prove, because
it only has to guarantee fractional progress. Once such a
guarantee is available, it can be lifted to a full exploration result by
applying it repeatedly.

\begin{figure}[ht]
    \centering
    \includegraphics[width=0.8\textwidth]{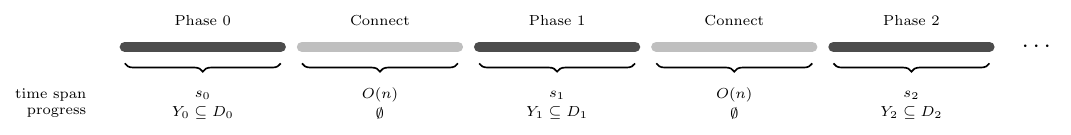}
\caption{
    Phase-based construction of a temporal walk that visits all vertices of
    a set $S\subseteq V$. In phase $i$, the walk visits some vertices $Y_i$ of the
    current set $D_i$ of unvisited vertices within some time span $s_i$.
    The connecting segments between phases are buffer walks which ensure that
    the phase walks can be concatenated to one final temporal walk.
}\label{fig:meta}
\end{figure}

We use a variant of this idea. Instead of always aiming
to visit all vertices of~$V$ quickly, we fix a target set $S\subseteq V$ and want to
visit all vertices of $S$. The set $S$ will be one $2$-WL
fiber. We build the temporal walk in phases; see \Cref{fig:meta}. At the
beginning of phase $i$, let $D_i\subseteq S$ be the set of vertices of $S$ that
have not yet been visited, with $D_0\coloneqq S$. During phase $i$, we
obtain a temporal walk $W_i$. Letting
$Y_i\subseteq D_i$ be the set of vertices of $D_i$ visited by $W_i$, we set $D_{i+1}\coloneqq D_i\setminus Y_i$. 

The temporal walks from the individual phases do not necessarily fit
together directly since a single-phase argument typically guarantees a walk with the
desired progress, but not one that starts at the current endpoint of the walk
constructed so far. We therefore leave a buffer of time steps between
consecutive phases. Since every snapshot is connected, any current vertex can
reach any other vertex within at most $n$ time steps~(\Cref{lem:reachability}). Thus, a buffer of length $n$ suffices to connect the
endpoint of one phase to the start vertex of the next phase. Concatenating the
phase walks with these buffer walks yields one temporal walk that visits all
vertices of $S$.



\subsection{From Automorphism Orbits to \texorpdfstring{$\boldsymbol{2}$-WL}{2-WL}}\label{ssec:twl}
Dogeas, Erlebach, Kammer, Meintrup, and Moses
Jr.~\cite{DogeasEKJMW24} showed that connected temporal graphs can be
explored in $O(rn^{1+\varepsilon})$ time steps, where $r$ is the number of
automorphism orbits. We recall the main construction, as it will help understand the intricacies that come with using restricted notions of symmetry.
As is common, their proof uses a phase-based construction as sketched above. 

In one phase, the goal is to make guaranteed progress in the exploration of a fixed automorphism orbit $S$. The key ingredient for their proof is that, for every fixed
$\varepsilon>0$, one can deterministically construct a temporal walk $W$ that spans
$O(r|S|^{1+\varepsilon})$ time steps and visits at least $\delta |S|$ vertices of $S$,
for some fixed constant $\delta\in(0,1)$. The walk $W$ need not visit many
vertices of the current unvisited set $D_i\subseteq S$; in the worst case, it
may make no progress in the current phase.
The orbit
structure is then used to repair it and ensure the required progress. Here is
where the full symmetry captured by automorphisms is used: apply vertex by
vertex an automorphism to $W$ to obtain a walk $W'$ that visits different
vertices of $S$. By an averaging
argument over the automorphism group, one can find an automorphism so that the
new walk visits many vertices of $D_i$.

To construct the walk $W$ that visits many vertices of $S$, the authors
of~\cite{DogeasEKJMW24} essentially use a divide-and-conquer construction.
Roughly speaking, the construction recursively builds two smaller walks
$\Wleft$ and $\Wright$ entirely independently
\new{(where $\ell$ and $r$ denote left and right, respectively)} such that $\Wright$ starts a
certain number of time steps after $\Wleft$, transforms $\Wright$ by an
automorphism, and finally connects them by an intermediate walk $\Wglue$. For
the choice of the start time of $\Wright$, it is important to note that we
cannot pick a time step when we can be sure that every vertex of $S$ is
reachable, as this could incur up to $n$ time steps at every recursive 
step~(\Cref{lem:reachability}) for each of the walks $\Wglue$. Thus, only the set of
vertices $X \subseteq S$ reachable from the endpoint of $\Wleft$ within a short
time span are valid start vertices for $\Wright$. But as the original $\Wright$
is constructed absolutely independently of $\Wleft$, it need not start at a
vertex of $X$, and it may visit many of the same vertices as $\Wleft$. The
automorphism machinery is used to repair both issues at once: one applies a
suitable automorphism $\sigma$ to $\Wright$ to obtain a walk $\Wright_{\sigma}$
so that its start vertex lies in the reachable set and the transformed walk has
small overlap with $\Wleft$. This is the key strength of the orbit-based
divide-and-conquer construction: the two recursive walks can be constructed
independently because automorphisms provide a repair mechanism in the merge
step. Constructing the described walks requires exhaustive search over
automorphisms or brute-force enumeration of temporal walks. Moreover, computing
automorphism orbits is only known to be possible in quasipolynomial time in
general via Babai's algorithm~\cite{Babai19}. 

Our intention is to reduce the computational complexity. To this end, we relax
the equivalence notion between vertices from being contained in the same orbit
to having the same $2$-WL color, i.e.,\ being in the same $2$-WL fiber. This
weaker notion still guarantees that equivalent vertices share certain
combinatorial properties. More precisely, the $2$-WL colors encode the number
of temporal walks between pairs of vertices. Hence, in particular, equivalent
vertices have equal counts of temporal walks into other fibers. Contrary to
automorphism orbits, the $2$-WL fibers are known to be computable in polynomial
time, so the computational issues described above disappear. In addition, since
the partition into $2$-WL fibers is coarser than the partition into orbits, we
obtain a better factor in our complexity analysis. However, and crucially, we
lose the repair mechanism that the perfect symmetry captured by the
automorphisms offers. Nonetheless, $2$-WL allows us
to, in a sense, emulate the repair step with a probabilistic
procedure, outlined later in this section.

Focus on a single phase $i$. Fix a $2$-WL fiber $S$, and let $D_i\subseteq S$ be the
vertices of $S$ that have not yet been visited. In this phase, a walk
that merely visits many vertices of $S$ is not enough: it has to visit many
unseen vertices, i.e.,\ many vertices of the current set $D_i$ since we cannot
repair our walks as in the orbit-based approach. We still build the phase walk
recursively, with a left part, a connecting part, and a right part. 

A recursive construction starts at some vertex $u$ of $S$ in time step $t_1$ and first
computes recursively a left walk $\Wleft$, starting at $u$ in time step $t_1$.
Suppose that $\Wleft$ ends at a vertex $v'$ at time step $t_2$. The algorithm
then considers the set $X_{v'}\subseteq S$ of vertices that can be reached by
temporal walks starting at $v'$ in time step $t_2$ and ending at a time step
$t_3$, where the choice of $t_3$ will be specified later. It chooses a vertex
$u'$ from $X_{v'}$, follows a \new{connecting temporal walk} $\Wglue$ from $v'$ to $u'$, and then
starts the right recursively computed walk $\Wright$ at $u'$ in time step
$t_3$. The walk returned by the construction is the concatenation of $\Wleft$, 
$\Wglue$, and $\Wright$. But the recursion can no longer be
viewed as two independently constructed sub-walks followed by a repair step.
Thus, the recursive choices themselves must ensure the properties needed in the
phase: the final walk makes progress on $D_i$, the left and right parts both
contribute relevant vertices, and their contributions do not overlap too much.
The right recursive construction must depend on the outcome of the left
recursive call, and the set $D_i$ has to be part of the analysis from the
beginning.

We handle this dependence by introducing randomness. Intuitively, the resulting process is similar to a random walk, but
instead of choosing a random neighbor at each time step, it chooses a vertex
from a reachable set within a prescribed time span, and then follows some
temporal walk to that vertex. In our
recursion, after $\Wleft$ has been constructed and ends at $v'$ at time step
$t_2$, we choose the start vertex of $\Wright$ uniformly at random (u.a.r.)
from the set $X_{v'}\subseteq S$, described above. The set $X_{v'}$ depends on
the actual end point of $\Wleft$, so the left and right parts are not
independent in the usual probabilistic sense. The role of $2$-WL is to show
that the resulting transition probabilities are nevertheless balanced enough to
make the recursive analysis go through.

\begin{figure}[ht]
    \centering
    \includegraphics[width=0.8\textwidth]{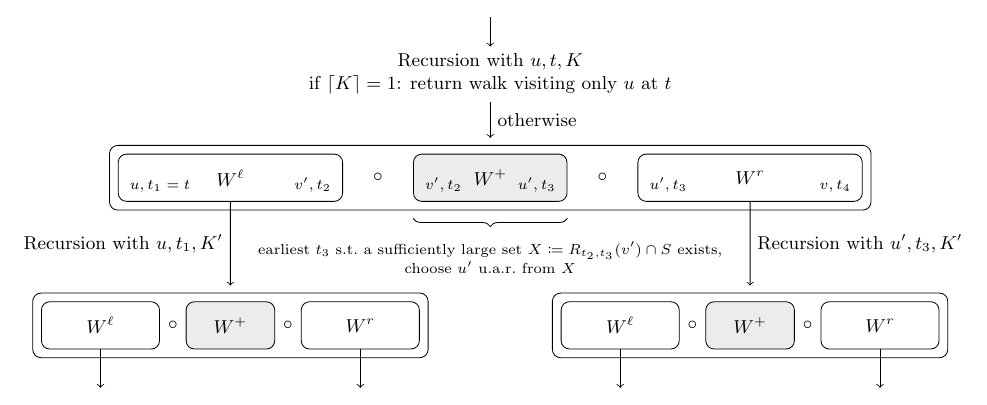}
\caption{
    Sketch of our randomized recursive procedure. The recursion is started with
    exploration scale~$K$, a start vertex $u$, and a start time $t$. If $\lceil K\rceil\ = 1$, we
    return the trivial temporal walk that stays at $u$ at time step $t$.
    Otherwise, we first construct a walk $\Wleft$ with a smaller exploration scale $K'$,
    starting at $u$ at time step $t_1=t$
    and ending at vertex $v'$ at time step $t_2$.
    Starting from $v'$ in time step $t_2$,
    we construct a sufficiently large set $X$ of reachable vertices.
    We then choose a vertex
    $u'$ u.a.r. from $X$, and use $u'$ as the start of the
    recursively constructed walk $\Wright$.
}
    \label{fig:recursive_sketch}
\end{figure}

The recursion is governed by a parameter $K$, which we call the
\emph{exploration scale}. Informally, and with some imprecision for the sake of
intuition, a construction with exploration scale $K$ 
constructs a walk that aims to visit $\Theta(K)$
distinct vertices of the fiber $S$.
The initial
recursive construction uses an exploration scale of order roughly $|S|$, so the goal is
for the recursively constructed walk 
to visit a substantial part of $S$.
Each recursive step then splits this task into a left and a right walk,
each of which should be thought of as continuing with exploration scale roughly
$K/2$. The recursion stops once $\lceil K\rceil=1$; in this case, the algorithm
returns a trivial temporal walk that visits one vertex in one time step. 
See \Cref{fig:recursive_sketch} for a schematic overview.

The difficulty is to prove that the randomized construction actually produces
a temporal walk
that visits many vertices of the current
set $D_i\subseteq S$ from the phase under consideration.
We show that the procedure constructs a temporal walk $W$ that
visits $\Omega(|D_i|/\log |S|)$ vertices of $D_i$ in expectation\journal{~(\Cref{lem:partial-exploration})}, while $W$ has time span
$O(q|S|\log |S|)$\journal{~(\Cref{lem:num-time-steps})}. In the next subsection, we give
an overview of the proof of the expected-progress bound, then sketch the
time-span analysis, and finally explain the derandomization.

\subsection{Randomized Partial Exploration}\label{ssec:rpe}
Before we can analyze the randomized recursion, we collect some useful structural properties of the
$2$-WL algorithm. Let $\chi^2\colon V^2\to C$ be the final $2$-WL coloring
of the temporal graph, obtained after encoding the time labels as edge colors and executing the algorithm on the edge-colored graph. If
$u,v,u',v'\in V$ satisfy $\chi^2(u,v)=\chi^2(u',v')$, then, for any time steps
$t,t'\in\{1, \ldots, L\}$, the number of temporal walks that start at $u$ at time step $t$
and end at $v$ at time step $t'$ equals the number of temporal walks
that start at $u'$ at time step $t$ and end at $v'$ at time step $t'$.

Now fix a fiber $S$ and two time steps $t\leq t'$. Consider the bipartite
graph $H$ whose two sides are copies of $S$, and where a vertex $u$ on the left
is adjacent to a vertex $v$ on the right if $v$ is reachable from $u$ between
time steps $t$ and $t'$. The $2$-WL property described above implies that $H$ is biregular:
all vertices on the left have the same degree, and all vertices on the
right have the same degree\journal{~(\Cref{lem:walk-regular-biregular})}. If we choose a vertex on the left u.a.r.\ and then choose an outgoing edge
u.a.r., we obtain a matrix $M\in\mathbb{R}^{S\times S}$ of transition probabilities between vertices where the row of
$M$ is indexed by the start vertex $u\in S$, and the column is indexed by the
endpoint $v\in S$ on the right. Since $H$ is biregular, it turns out that this matrix is
\emph{doubly stochastic}, meaning that each row and each column sums to $1$\journal{~(\Cref{lem:doubly})}.  
This models the transition that occurs in our recursion~(\Cref{fig:recursive_sketch}) when we choose the start vertex of
$\Wright$ uniformly from a reachable set. Thus, whenever the endpoint of
$\Wleft$ is uniformly distributed over~$S$, the start vertex of $\Wright$
remains uniformly distributed over~$S$. Since products of doubly stochastic
matrices are doubly stochastic themselves, we can show that the same balancedness persists through
the recursive construction.

To analyze the procedure, we turn to the recursion tree $\cT$ that represents the relations between the recursive calls. 
Each node $x$ of $\cT$ represents the recursive computation
to construct a walk $W_x$.
If $x$ is an internal node, then $W_x$
is composed of three sub walks: a left walk $\Wleft_x$, computed
recursively, a
connecting walk $\Wglue_x$, and a right walk $\Wright_x$, also computed recursively.
If $x$ is a leaf, then $x$ represents the base case of the recursion and $W_x$
is the trivial temporal walk that stays at one vertex of $S$ for one time step.

We write $\cT_x$ for the subtree of $\cT$ rooted at $x$. The walk $W_x$ is produced by the full recursive computation inside $\cT_x$. If $x$ is an
internal node with children $x_{\ell}$ and $x_r$, then $x_{\ell}$ represents
the recursive construction that computes $\Wleft_x$, and $x_r$ represents the recursive
construction that computes $\Wright_x$. The node $x$ itself additionally computes the
connecting walk $\Wglue_x$ between them. 
Let $x$ be a node of $\cT$. We say that $x$ \emph{starts at} $u\in S$ if the
recursive construction represented by $x$ is invoked with start vertex $u$,
equivalently, if $W_x$ starts at $u$. We say that $x$ \emph{ends at} $v\in S$
if $W_x$ ends at $v$. If $x$ is a leaf, then the two vertices are the same:
the walk $W_x$ is the trivial walk that stays at its start vertex for
one time step. Finally, we say that $x$ \emph{visits} a vertex $u\in S$ if
$W_x$ contains $u$.

For the analysis, we model the recursive procedure by transition matrices,
outlined next.
Let $x$ be an internal node of $\cT$ with left child $x_\ell$ and right child $x_r$.
Suppose that the walk corresponding to $x_\ell$ ends at vertex $v'$ at time
step $t_2$.
We choose the start vertex of the right recursive
construction uniformly from a reachable set
$X_{v'}=R_{t_2,t_3}(v')\cap S$, where $t_3$ is chosen sufficiently large
for our analysis to go through.
We define
the \emph{internal transition matrix} $P_x\in\mathbb{R}_{\geq 0}^{S\times S}$ by
\[
    (P_x)_{v',u'} =
    \begin{cases}
        1/|X_{v'}|, & \text{if } u'\in X_{v'},\\
        0,       & \text{otherwise.}
    \end{cases}
\]
Thus, $P_x$ models the random transition from the endpoint of $\Wleft_x$ to
the start vertex of $\Wright_x$.

We also define \emph{endpoint-transition matrices}. For every node $x$ of
$\cT$, let $E_x\in\mathbb{R}_{\geq 0}^{S\times S}$ be given by
\[
    (E_x)_{u,v}
    \coloneqq
    \Pr[x\text{ ends at }v \mid x\text{ starts at }u].
\]
If $x$ is a leaf, then $E_x$ is the identity matrix. If $x$ is an internal node
with children $x_\ell$ and $x_r$, then $E_x = E_{x_\ell} P_x E_{x_r}$. This is because the walk $W_x$ first follows the walk $\Wleft_x$,
then makes the random internal transition described by $P_x$
to the start vertex of $\Wright_x$, and
finally follows the walk $\Wright_x$.

Finally, for every node $x$ and every leaf $y$ in
$\cT_x$, we define the \emph{leaf-visit matrix}
$L_{x,y}\in\mathbb{R}_{\geq 0}^{S\times S}$ by
\[
    (L_{x,y})_{u,v}
    \coloneqq
    \Pr[y\text{ visits }v \mid x\text{ starts at }u].
\]
If $x=y$ is a leaf, then $L_{x,y}$ is the identity matrix. If $x$ is an
internal node and $y$ lies in the left subtree $\cT_{x_{\ell}}$ of $\cT_x$, then
$L_{x,y}=L_{x_\ell,y}$.
If $y$ lies in the right subtree $\cT_{x_{r}}$ of $\cT_x$, then
$L_{x,y}=E_{x_\ell}P_xL_{x_r,y}$.
    This is because the start vertex of
$x_r$ is determined by first following the walk $\Wleft_x$
and then applying the internal transition $P_x$; subsequently, the vertex
visited by
$y$ is described by $L_{x_r,y}$.

The biregularity statement above implies that each matrix $P_x$ is in fact doubly
stochastic\journal{~(\Cref{lem:doubly})}. It follows
inductively that each $E_x$ and each $L_{x,y}$ is doubly stochastic as well\journal{~(\Cref{lem:recursive-transition-matrices})}.

\paragraph*{Lower Bound on the Number of Distinct Vertices}
We now sketch the lower bound on the number of distinct vertices of
$D\subseteq S$ visited in one phase. Let $\cT$ be the recursion tree of our
recursive procedure, let $x_0$ be its root, and let $K_0$ be the exploration
scale used by the recursive construction represented by $x_0$.
Recall that, for a leaf node $x$ of $\cT$, the walk $W_x$ is trivial and
visits a single vertex $u$ of $S$; we call this vertex $u$ the \emph{leaf visit} of
$x$. Let $Z_D$ be the number of leaf nodes $x$ whose leaf visit is a vertex of
$D$, counted with multiplicity. We show that, for every fixed leaf node $x$ of
$\cT$, the leaf visit of $x$ is uniformly distributed over $S$. 
This implies that each leaf contributes $|D|/|S|$ in expectation to $Z_D$.
Together with the fact that $\cT$ has $\Theta(K_0)$ leaves\journal{~(\Cref{lem:number-of-leaves})}, we can show that
$\bE[Z_D]=\Omega(K_0|D|/|S|)$\journal{~(\Cref{lem:expected-leaf-visits})}. This bound only counts visits with
multiplicity and therefore does not yield any lower bound on the number of
distinct vertices visited. In particular, $Z_D$ may be
large even if the same vertex of $D$ is visited over and over again. To pass
from visits with multiplicity to distinct vertices, we count collisions.
We define a \emph{collision} as an unordered pair of leaves $\{y, z\}$
    of $\cT$
such that $y$ visits the same vertex $u$ as $z$ and $u \in D$.
Let $Q_D$ be the number of collisions.
$Z_D-Q_D$ is a lower bound on the number of distinct
vertices of $D$ visited. Thus, it remains to prove that
$\bE[Z_D-Q_D]=\Omega(|D|/\log |S|)$.

\begin{figure}[ht!] \centering \includegraphics[width=0.8\textwidth]{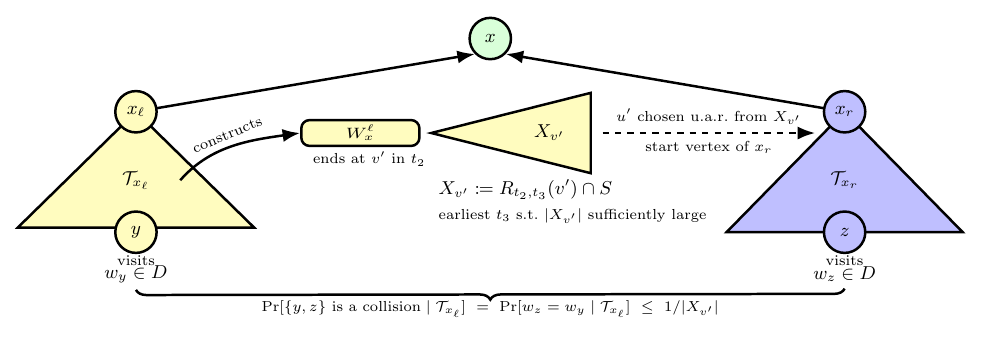}
\caption{
Sketch of the walks in the computation represented by a node $x$ and the
resulting probability bound on collisions. Everything colored in yellow is
considered precomputed and hence fixed. We condition on the entire left subtree
$\cT_{x_\ell}$. Then, for every leaf $y$ of $\cT_{x_\ell}$, the vertex $w_y\in
D$ visited by $y$ is fixed. Moreover, the walk $\Wleft_x$ ends at a fixed
vertex $v'$ at a fixed time step $t_2$, and therefore the set $X_{v'} \coloneq
R_{t_2,t_3}(v')\cap S$ is fixed as well, where $t_3$ is chosen such that
$|X_{v'}|$ is sufficiently large. The start vertex of the right recursive
construction is chosen uniformly at random from $X_{v'}$. Since the transition
matrices in the right subtree are doubly stochastic, for a fixed leaf $z$ of
$\cT_{x_r}$ and any fixed vertex $w\in D$, the probability that $z$ visits $w$
is at most $1/|X_{v'}|$. 
For every fixed pair of leaves $y\in
\cT_{x_\ell}$ and $z\in \cT_{x_r}$, the probability that $y$ and $z$ visit the
same vertex is at most~$1/|X_{v'}|$.
}
\label{fig:coinciding}
\end{figure}

The delicate part is bounding $Q_D$, because the two sides of a recursive call
are not independent: the start vertex of the right recursive walk is chosen
only after the left walk has ended, and the set $X$ from which it is chosen
depends on the endpoint of the left walk. Every collision $\{y,z\}$ can
be assigned to the lowest common ancestor $x$ of $y$ and $z$. Fix such a node
$x$ with left child $x_{\ell}$
and assume the walk $\Wleft_x$ has been computed.
Then the
leaf visits in $\cT_{x_{\ell}}$ are fixed, the walk $\Wleft_x$ ends at some fixed vertex $v$, and
the reachable set $X_v\subseteq S$ for the right recursive construction is fixed as
well.
The right recursive computation
for $\Wright_x$ starts from a vertex of $X_v$ chosen u.a.r.
Hence, every vertex of $S$ has probability at most $1/|X_v|$ of being 
chosen as the start
vertex of $\Wright_x$. The later leaf-visit distributions in the right subtree are obtained by
multiplying with doubly stochastic matrices, which cannot increase the maximum
probability assigned to any vertex\journal{~(\Cref{obs:product-ds-row})}. Consequently,
for every leaf $z$ in the right subtree and every fixed vertex $w\in S$, the
probability that $z$ visits $w$ is at most~$1/|X_v|$. See \Cref{fig:coinciding}
for a sketch.

By choosing the sizes of the sets $X_v$ carefully, we can ensure that the total
expected contribution of all collisions is only a constant fraction of
$\bE[Z_D]$\journal{~(\Cref{lem:coinciding-leaves})}. 
This gives the required upper bound
on $\bE[Q_D]$, hence a lower bound on $\bE[Z_D-Q_D]$,
which yields that,
for any $D \subseteq S$,
the walk computed by our entire recursive construction
visits many vertices of $D$ in expectation\journal{~(\Cref{lem:partial-exploration})}.

\paragraph*{Upper Bound on the Number of Time Steps} 
So far,
we have shown that the walk computed by our recursive construction
visits many different vertices of any set $D \subseteq S$
in expectation.
We have not yet considered the time span of the recursively constructed
walk. The time-span analysis relies on two ingredients. First, we carefully
balance, at each recursion level, the number of time steps that the glue walks
are allowed to span with the size of the reachable sets from which the right
recursive construction may start. Second, we choose the initial recursion parameter as
$K_0=\Theta(|S|/\log |S|)$. This choice allows us to bound the number of leaves
of the recursion tree by $O(|S|/\log |S|)$. With the bound on the time
spent at the internal nodes of the recursion tree, this yields a total time
span of $O(q|S|\log |S|)$ for the constructed temporal walk\journal{~(\Cref{lem:num-time-steps})}.

    To explore the entire set $S$, we then use the phase strategy described at the
beginning of the section and visualized in \Cref{fig:meta}. In phase $i$, the
current set of unvisited vertices is $D_i\subseteq S$. The partial-exploration
guarantee\journal{~(\Cref{lem:partial-exploration})} gives a temporal walk that visits
$\Omega(|D_i|/\log |S|)$ vertices of $D_i$ and has time span
$O(q|S|\log |S|)$. 
Hence, each phase removes an $\Omega(1/\log |S|)$ fraction of the remaining
vertices. Intuitively, as long as at least $|S|/2$ vertices remain unvisited,
each phase visits roughly $|S|/(2\log |S|)$ new vertices, so after
$O(\log |S|)$ phases approximately half of $S$ has been visited. 
Then, as long as at least $|S|/4$ vertices remain unvisited, each phase still
visits roughly $|S|/(4\log |S|)$ new vertices. Continuing in this way, the
number of unvisited vertices decreases by a constant factor every
$O(\log |S|)$ phases.
By this argument, we can then show that after $O(\log^2 |S|)$ phases, only
$O(\log^2 |S|)$ vertices of $S$ remain unvisited. Each remaining vertex can be
visited in $O(n)$ time steps~(\Cref{lem:reachability}), contributing $O(n\log^2
|S|)$ additional time steps in total. Thus, we can explore $S$ in $O(q|S|\log^3
|S|+n\log^2 |S|)$ time steps.

\paragraph*{Derandomization}
The preceding argument is only existential: it shows that some outcome of the
randomized construction gives sufficient progress in each phase, but it does
not yet give a deterministic algorithm for finding such an outcome. 
Surprisingly, we can turn this
existential proof into a deterministic polynomial-time algorithm by derandomizing the
construction via the method of conditional expectations; see, e.g.,
\cite[Chapter~16.1]{AlonS00}.

Recall that the analysis considers $\bE[Z_D-Q_D]$, where $Z_D$ counts
leaf visits in $D$ and $Q_D$ counts collisions in $D$. In the exploration
algorithm, the set $D$ is the current set of vertices of $S$ that
have not yet been visited. For a partial sequence $\sigma$ of fixed random
choices, we consider the conditional expectation
$\bE[Z_D-Q_D\mid\sigma]$. Initially, for $\sigma=\emptyset$, this is the
expectation analyzed above.

We fix the random choices one by one. Suppose that $\sigma$ is the sequence of
choices fixed so far. For the next random choice, we iterate over all possible
outcomes $c$ and compute
$\bE[Z_D-Q_D\mid\sigma_c]$, where $\sigma_c$ is the sequence obtained
from $\sigma$ by appending $c$. We then choose an outcome $c$ maximizing this
conditional expectation and replace $\sigma$ with $\sigma_c$. This cannot
decrease the conditional expectation, because
$\bE[Z_D-Q_D\mid\sigma]$ is the average of
$\bE[Z_D-Q_D\mid\sigma_c]$ over all possible outcomes $c$ of the next
random choice. Repeating this until all random choices have been fixed yields a
deterministic temporal walk whose value of $Z_D-Q_D$ is at least the original
lower bound on $\bE[Z_D-Q_D]$.
It just remains to argue that this procedure runs in polynomial time. For this,
we show that there are only polynomially many choices to consider\journal{~(\Cref{cl:numchoices})} and that each required conditional expectation can be
evaluated in polynomial time\journal{~(\Cref{cl:evaluation})}.

\subsection{Temporal Rendezvous}\label{ssec:trp}
The rendezvous result follows
from the single-fiber exploration bound~(\Cref{thm:one-part-exploration}). Recall that, in $\trp$, two agents $a_1$ and
$a_2$ know the temporal graph, but do not share a common vertex labeling. To
prove \Cref{thm:rendezvous}, we describe a strategy that both agents can
compute in polynomial time such that the agents meet in $O(n \log^3 n)$ time
steps. Both agents compute the $2$-WL fibers and choose a smallest fiber $S$,
with ties broken canonically by the $2$-WL colors. 
Since the $2$-WL coloring is canonical, both agents choose the same fiber.
Agent $a_1$ moves to a vertex of $S$ within at most $n$
time steps and waits there. Agent $a_2$ waits for $n$ time steps and then
explores all vertices of $S$ using \Cref{thm:one-part-exploration}, or until
agent $a_2$ finds agent $a_1$. Since $S$ is a smallest fiber, we have $|S|\leq
\lceil n/q\rceil$ and thus agent $a_2$ explores all of $S$ within $O(q|S|
\log^3 |S|+n\log^2 n)=O(n \log^3 n)$ time steps. It remains to note that the
strategy can be implemented in polynomial time. The agents must first compute the underlying graph together with the
    edge labels representing the time steps each edge is active at, which can be
    done in linear time with a scan over all snapshots. The $2$-WL fibers can
    then be computed in time $O(n^3\log n)$~\cite{ImmermanL90,ImmermanS19}.
    \new{This running-time bound applies after preprocessing the temporal
    edge labels into integer colors; in particular, it holds for edge-colored graphs.}
    \Cref{thm:one-part-exploration} gives a polynomial-time algorithm for agent
    $a_2$ to explore all of $S$, and the temporal walks used to reach the fiber $S$
    can be computed efficiently using standard algorithms for temporal
reachability~\cite{WuCYSYW16}.

\journal{ 
\section{Preliminaries}\label{sec:preliminaries}
\preliminaries

\section{\texorpdfstring{The Connection Between $\boldsymbol{2}$-WL and Temporal-Walk Counts}{The Connection Between 2-WL and Temporal-Walk Counts}}\label{sec:twowl}
For the remainder of this section, let $\deftg$ be a temporal graph with vertex
set $V$ and lifetime $L$. Let $\cQ$ be the partition of $V$ into its $2$-WL
fibers, and let $\chi^2\colon V^2\to C$ be the $2$-WL coloring inducing $\cQ$
, that is, two vertices $u,v\in V$ belong to the same fiber if and
only if $\chi^2(u,u)=\chi^2(v,v)$.

Recall that $\chi^2$ refines the initial coloring $\chi^2_0$, which itself
encodes the adjacency and equality type of the vertex pairs. Hence, for
vertices $u,v,w$ with $v \neq w$, we always have $\chi^2(u,u) \neq \chi^2(v,w)$,
that is, fibers are indeed color classes of the entire set $V^2$.

The following two lemmas state well-known properties of $2$-WL. We include proofs for completeness.

\begin{lemma}\label{lem:vertexedge}
  For all fibers $S_1,S_2,S,S'\in\cQ$ with $S_1 \neq S_2$, and for $u_1 \in S_1$, $u_2 \in S_2$, $v \in S$, $v' \in S'$, 
  \[
    \chi^2(u_1,v) \neq \chi^2(u_2,v') \quad \text{and} \quad \chi^2(v,u_1) \neq \chi^2(v',u_2).
  \]
\end{lemma}

\prooflemvertexedge

\begin{lemma}\label{lem:twlcounts}
For all fibers $S,S'\in\cQ$, all vertices $u,v\in S$, and every color
$\gamma\in C$, it holds that
\[
    |\{x\in S'\mid \chi^2(u,x)=\gamma\}|
    =
    |\{x\in S'\mid \chi^2(v,x)=\gamma\}|.
\]
Analogously,
\[
    |\{x\in S'\mid \chi^2(x,u)=\gamma\}|
    =
    |\{x\in S'\mid \chi^2(x,v)=\gamma\}|.
\]
\end{lemma}
\prooflemtwlcounts

\begin{lemma}\label{lem:twlwalks}
For all pairs $(u,v),(u',v')\in V^2$ with
$\chi^2(u,v)=\chi^2(u',v')$ and all $t \leq t'\in[L]$,
the number of
temporal walks from $u$ to $v$ starting at time step $t$ and ending at time
step $t'$ is equal to the number of temporal walks from $u'$ to $v'$
starting at time step $t$ and ending at time step $t'$.
\end{lemma}
\prooflemtwlwalks

For fibers $S,S'\in\cQ$ and time steps $t,t'\in[L]$ with $t\leq t'$, let
$H_{t,t'}(S,S')$ be the complete bipartite weighted graph with bipartition
$(S,S')$ in which the weight of the edge between $u\in S$ and $v\in S'$ is the
number of temporal walks starting at time step $t$ in $u$ and ending at
time step $t'$ in $v$. The \emph{weighted degree} of a vertex is the sum of the
weights of its incident edges. We call a weighted bipartite graph
\emph{biregular} if, on each side of the bipartition, all vertices have the same
weighted degree (which may, however, be different between the sides).
We denote by $\operatorname{supp}(H_{t,t'}(S,S'))$ the \emph{support graph} of
$H_{t,t'}(S,S')$, defined as the bipartite graph on $(S,S')$ that contains an
edge between $u\in S$ and $v\in S'$ if and only if the corresponding weight in
$H_{t,t'}(S,S')$ is at least $1$. The notions apply also for unweighted
bipartite graphs by considering them as graphs with uniform edge weight~$1$.

\begin{lemma}\label{lem:walk-regular-biregular} Let $S,S'\in\cQ$ and let
$t \leq t'\in[L]$. Then $H_{t,t'}(S,S')$ and
$\operatorname{supp}(H_{t,t'}(S,S'))$ are biregular.
\end{lemma}
\prooflemwalkregularbiregular

The next observation follows
from the fact that every vertex can reach
itself in one time step.
\begin{observation}\label{obs:positive-degree-same-fiber}
    Let $S\in\cQ$ and let $t\leq t'\in[L]$. Then every vertex of
    $H_{t,t'}(S,S)$ has weighted degree at least $1$, and every vertex of
    $\operatorname{supp}(H_{t,t'}(S,S))$ has degree at least $1$.
\end{observation}

A matrix $M\in\mathbb{R}_{\geq 0}^{n\times n}$ is \emph{doubly stochastic} if
every row of $M$ sums to $1$ and every column of $M$ sums to $1$. 
We use the
following two elementary observations.

\begin{observation}
\label{obs:product-ds}
Let $M,M'\in\mathbb{R}_{\geq 0}^{n\times n}$ be doubly stochastic matrices,
for some integer $n\geq 1$. Then $M \cdot M'$ is doubly stochastic.
\end{observation}
\begin{observation}

\label{obs:product-ds-row}
Let $M\in\mathbb{R}_{\geq 0}^{n\times n}$ be a doubly stochastic matrix and
let $\rho\in\mathbb{R}_{\geq 0}^{1\times n}$ be a row vector such that
every entry of $\rho$ is at most $\alpha$, for some $\alpha\geq 0$. Then
every entry of $\rho M$ is at most $\alpha$.
\end{observation}

In the next section, we analyze a procedure that repeatedly chooses a vertex
uniformly at random from a set of possible endpoints of temporal walks. Such a
choice gives a transition matrix $P$, where $P_{u,v}$ is the probability of
choosing $v$ when starting from $u$. 
The support graph defined above induces doubly stochastic
transition matrices, as shown next.
We make use of \Cref{obs:positive-degree-same-fiber} to ensure the degrees
of vertices in the
bipartite graphs considered in the upcoming lemma is not zero.

\begin{lemma}\label{lem:doubly}
    Let $S\in\cQ$, and let $t\leq t'\in[L]$. Let $H\coloneq
    \operatorname{supp}(H_{t,t'}(S,S))$, and let $d\geq 1$ be the common degree of
    the vertices in the left copy of $S$ in $H$. Define $P\in\mathbb{R}_{\geq
    0}^{S\times S}$ by setting $P_{u,v}\coloneq 1/d$ if $(u,v)$ is an arc of
    $H$, and $P_{u,v}\coloneq 0$ otherwise. Then $P$ is doubly stochastic.
\end{lemma}

\prooflemdoubly

\section{Temporal Exploration via 2-WL Fibers}\label{sec:algorithm}
For the remainder of this section, let $\deftg$ be a connected temporal graph
with vertex set $V$ and lifetime $L$, and let $\cQ$ be the partition of $V$
into its $2$-WL fibers. In this section, we prove the algorithmic results of
the paper. The main part is the proof of \Cref{thm:one-part-exploration}: for a
fixed fiber $S\in\cQ$, we construct a temporal walk that visits all vertices of
$S$ within the claimed time span. Applying this single-fiber result to all
fibers one after another then yields \Cref{thm:exploration}. The implications
for $\trp$ have been shown in \Cref{ssec:trp}. To avoid irrelevant edge
cases, we handle small fibers separately. Throughout this section, whenever we
consider a fiber $S$, we assume that $|S|$ is larger than a sufficiently large
constant, otherwise, we visit all vertices of
$S$ directly by repeated applications of \Cref{lem:reachability}. This takes
$O(|S|n)=O(n)$ time steps and is therefore within any claimed bound below.

The proof proceeds in three parts. Recall that the general idea is to
repeatedly make progress on a set $D\subseteq S$ of vertices that have not yet
been visited. First, we give a randomized recursive procedure that is
independent of $D$. We then analyze the time span of the computed walk $W$ and show that, for every
fixed set $D\subseteq S$, the walk $W$ visits sufficiently many distinct
vertices of $D$ in expectation. Finally, we derandomize the construction using
conditional expectations and apply the resulting deterministic
partial-exploration procedure repeatedly until all vertices of $S$ have been
visited.

\subsection{Outline of the Procedure}\label{ssec:outline}
The procedure $\textsc{PartialExplore}(S,t)$ (\Cref{alg:sample}) constructs a temporal
walk that starts at a u.a.r.\ chosen vertex in $S$ at time step $t\in [L]$. For
convenience, we define the actual recursion with the helper procedure
$\textsc{Recurse}(u,t,K)$ that takes a fixed vertex of $S$ as the starting
vertex and an additional parameter $K$, which we call the \emph{exploration
scale}. One of our goals is to show that the recursion tree generated by a call
$\textsc{Recurse}(u,t,K)$ has $\Theta(K)$ leaves. We first fix some parameters.
Take $c\coloneq C\log |S|$ for a sufficiently large constant $C$ and let
$f\coloneq 1+(c-1)/c$. The first call of $\textsc{Recurse}$ uses the \emph{initial
exploration scale $K_0\coloneq |S|/c$}. If a call $\textsc{Recurse}(u,t,K)$ has
exploration scale $K$ larger than the base-case threshold ($\lceil K \rceil >
1$), it recursively constructs a left and a right walk, both with exploration
scale $K/f$. Between these two walks, it constructs a connecting segment that
provides at least $cK/f$ possible start vertices for the right walk, from which
one is chosen u.a.r. At the bottom of the recursion, i.e., at the leaves of the
recursion tree, the procedure returns a trivial temporal walk visiting a single
vertex of $S$ in one time step. While we are later interested in visiting many
vertices of a set $D\subseteq S$ of desirable vertices, the set $D$ is not
given to the procedure as an input. Instead, the random construction is
sufficiently balanced that, for every fixed set $D\subseteq S$, a temporal walk
returned by an invocation of $\textsc{Recurse}$ with exploration scale $K$
visits $\Omega(K\cdot |D|/|S|)$ distinct vertices of $D$ in expectation.
Applied to the outermost invocation with $K=K_0$, we obtain the desired bound
$\Omega(|D|/\log |S|)$. See also \Cref{alg:sample}.

\begin{algorithm}[ht!]
\caption{\textsc{PartialExplore}$(S,t)$}
\label{alg:sample}
\begin{algorithmic}
\Require A fiber $S$, a start time $t$
\Ensure A sampled temporal walk whose first vertex is chosen u.a.r.\ from $S$
and which starts at time step $t$
\State Choose a sufficiently large constant $C$
\State $c\coloneq C\log |S|$
\State $f\coloneq 1+(c-1)/c$
\Procedure{Recurse}{$u,t_1,K$}
    \If{$\lceil K\rceil=1$}
        \State \Return the temporal walk that stays at $u$ at time step $t_1$
    \EndIf
    \State $\Wleft\leftarrow\textsc{Recurse}(u,t_1,K/f)$
    \State $v'\leftarrow$ final vertex of $\Wleft$
    \State $t_2\leftarrow$ ending time of $\Wleft$
    \State $t_3\leftarrow\min\{t'\geq t_2\mid |R_{t_2,t'}(v')\cap S|\geq cK/f\}$
    \State $X_{v'}\leftarrow R_{t_2,t_3}(v')\cap S$
    \State Choose $u'$ u.a.r.\ from $X_{v'}$
    \State $\Wglue\leftarrow$ a temporal walk from $v'$ to $u'$
        starting at $t_2$ and ending at $t_3$
    \State $\Wright\leftarrow\textsc{Recurse}(u',t_3,K/f)$
    \State \Return $\Wleft\circ\Wglue\circ\Wright$
\EndProcedure
\State Choose $u\in S$ u.a.r.
\State $K_0\coloneq |S|/c$
\State \Return $\textsc{Recurse}(u,t,K_0)$
\end{algorithmic}
\end{algorithm}

Let $\cT$ be the \emph{recursion tree} of an invocation of
$\textsc{PartialExplore}(S,t)$, defined with the following notation. We use the
word \emph{node} only for nodes of $\cT$, in order to distinguish them from
vertices of the temporal graph. The root of $\cT$ is denoted by $x_0$ and
represents the outermost recursive call, i.e., 
$x_0$ represents the call $\textsc{Recurse}(u,t,K_0)$, where $u$ is
chosen u.a.r. from $S$. For a node $x$ of $\cT$, we write $K_x$
for the
exploration scale of the call represented by $x$, and we write $W_x$ for the
temporal walk returned by this call. For a node
$x$ of $\cT$, we define $\cT_x$ as the subtree of $\cT$ rooted at $x$. The walk
$W_x$ is the walk computed by the full recursive computation
represented by the subtree $\cT_x$. We therefore sometimes use the node $x$ and
the walk $W_x$ interchangeably when the meaning is clear. Every non-leaf node
$x$ has two children, $x_{\ell}$ and $x_{r}$, representing its left and right
recursive calls of $\textsc{Recurse}$ with exploration scales $K_{x_\ell}=K_{x_r}=K_x/f$. We
denote the walks returned by these two subcalls by $\Wleft_x$ and $\Wright_x$,
respectively, and we denote by $\Wglue_x$ the connecting walk computed at node
$x$. Thus, $W_x = \Wleft_x \circ \Wglue_x \circ \Wright_x$. Every leaf $x$
represents a base-case call of $\textsc{Recurse}$ with exploration scale satisfying $\lceil
K_x\rceil=1$; then $W_x$ is the trivial temporal walk that stays at
its start vertex for one time step. For vertices $u,v,w\in S$ and a node $x$ of
$\cT$, we say that $x$ \emph{starts at $u$} if $u$ is the first vertex of
$W_x$, we say $x$ \emph{ends at $v$} if $v$ is the last vertex of $W_x$, and we
say \emph{$x$ visits $w$} if $W_x$ contains the vertex $w$.

We define the \emph{shape} of $\cT$ to consist of its number of
leaves, the number of nodes at each depth, and, for every node $x$ of $\cT$,
the exploration scale $K_x$ and the start and end times of the walk $W_x$.
We show that the shape of $\cT$ depends
only on the initial exploration scale $K_0$ and the initial start time $t$. In
particular, it is independent of the vertices and temporal walks sampled during
the execution of the algorithm, formalized in
\Cref{lem:common-times} and \Cref{lem:number-of-leaves}.

We introduce a small auxiliary lemma
that formalizes basic properties of the recursion tree produced
by \Cref{alg:sample}, followed by a lemma
bounding the number of leaves of any recursion tree.
This lets us upper-bound the number of time steps
that the
walks constructed by \Cref{alg:sample} span.

\begin{lemma}
    \label{lem:common-times}
    Let
    $\cT$ be a recursion tree.
    Then the following properties are independent
    of all random choices made by the algorithm
    and depend only on the initial exploration scale $K_0$ and initial start time $t$.
    \begin{enumerate}
        \item 
            All nodes $x, x'$ at the same depth $i$ of $\cT$ have the
            same exploration scale $K_x=K_{x'}=K_0/f^i$ with $f\coloneq 1+(c-1)/c$.

        \item $\cT$ is a binary tree with height $h=\Theta(\log K_0)$
            where all leaves have depth $d=h$.

        \item For every node $x$ of $\cT$, the start time and end
            time of the walk returned by the recursive call represented by
            $x$ are fixed.
    \end{enumerate}
\end{lemma}
\prooflemcommontimes

The previous lemma only gives a polynomial bound on the number of leaves in
terms of the initial exploration scale $K_0$. However, we need a linear bound.
This is shown in the next lemma, which is then used to bound the time span of
walks returned by \Cref{alg:sample}. For both upcoming proofs, we use a small
intermediate claim.

\begin{claim}\label{clm:boundf}
    Let $f\coloneq 1+(c-1)/c$ for $c=C\log |S|$, where $C>1$ is a sufficiently large
    constant and $|S|\geq 2$. Then $(2/f)^d=O(1)$ for every
    $d=O(\log |S|)$.
\end{claim}
\proofclboundf

\begin{lemma}\label{lem:number-of-leaves} 
    Let $\cT$ be a recursion tree and let $x$ be an internal node of
    $\cT$ with exploration scale $K_x$. The subtree $\cT_x$ has
    $\Theta(K_x)$ leaves. 
\end{lemma}
\prooflemnumberofleaves

We can now upper bound the time span of walks returned by
\textsc{PartialExplore}$(S,t)$, that is, by the initial call to
\textsc{Recurse} made in \Cref{alg:sample}.

\begin{lemma}\label{lem:num-time-steps}
Any walk returned by \Cref{alg:sample}
spans $O(q|S|\log |S|)$ time steps.
\end{lemma}
\prooflemnumtimesteps

\subsection{Analysis of the Procedure}\label{ssec:analysis}
We now analyze the algorithm probabilistically, with the goal of lower bounding
the expected number of vertices visited in an arbitrary set of desirable
vertices $D \subseteq S$. We use the matrix notation introduced in
\Cref{ssec:rpe}; for convenience, we recall it here.

We first define the transition matrix for the random choice made at an
internal node. Let $x$ be an internal node of $\cT$, with left and right
children $x_\ell$ and $x_r$, respectively. Suppose that the left walk
$\Wleft_x$, equivalently the walk $W_{x_\ell}$, ends at the vertex $v'$. The
algorithm then chooses the start vertex of the right recursive call uniformly
from the set $X_{v'}\subseteq S$. This choice is described by the
\emph{internal transition matrix} $P_x\in\mathbb{R}_{\geq 0}^{S\times S}$,
where
\[
    (P_x)_{v',u'} =
    \begin{cases}
        1/|X_{v'}|, & \text{if } u'\in X_{v'},\\
        0,          & \text{otherwise.}
    \end{cases}
\]
Thus, $P_x$ is defined only for internal nodes $x$; its row is indexed by the
endpoint $v'$ of the left walk, and its column is indexed by the chosen start
vertex $u'$ of the right walk.

For every node $x$ of $\cT$, we define the \emph{endpoint-transition matrix}
$E_x\in\mathbb{R}_{\geq 0}^{S\times S}$ by
\[
    (E_x)_{u,v}
    \coloneqq
    \Pr[x\text{ ends at }v \mid x\text{ starts at }u].
\]
If $x$ is a leaf, then $E_x$ is the identity matrix and $W_x$ is the
trivial temporal walk that stays at its start vertex for one time step. If $x$
is an internal node with children $x_\ell$ and $x_r$, then
\[
    E_x = E_{x_\ell} P_x E_{x_r}.
\]
This is because, conditioned on the start vertex of $x$, the walk first follows the left
recursive walk, then makes the random internal transition described by $P_x$,
and finally follows the right recursive walk. Finally, for every node $x$ of $\cT$ and every leaf node $y$ in
$\cT_x$, we define the \emph{leaf-visit matrix}
$L_{x,y}\in\mathbb{R}_{\geq 0}^{S\times S}$ by
\[
    (L_{x,y})_{u,v}
    \coloneqq
    \Pr[y\text{ visits }v \mid x\text{ starts at }u].
\]
If $x=y$, then $x$ is a leaf and $L_{x,y}$ is the identity matrix, because the
leaf visit is exactly the start vertex of the trivial walk $W_x$. Now suppose
that $x$ is an internal node with children $x_\ell$ and $x_r$. If $y$ lies in
the left subtree of $x$, then
\[
    L_{x,y}=L_{x_\ell,y}.
\]
This holds because, when $x$ starts at $u$, the left child $x_\ell$ also starts
at $u$. If $y$ lies in the right subtree of $x$, then
\[
    L_{x,y}=E_{x_\ell}P_xL_{x_r,y}
\]
and the distribution of the start vertex of $x_r$ is obtained by
first following the left recursive walk $\Wleft_x$ and then applying the
internal transition $P_x$; from that point on, the visit of $y$ is described by
$L_{x_r,y}$. Equivalently, if $\Wleft_x$ ends at a vertex $v'$, then the start
vertex $u'$ of $x_r$ is chosen according to the row $(P_x)_{v',\cdot}$, and the
distribution of the leaf visit of $y$ is then given by the row
$(L_{x_r,y})_{u',\cdot}$.

The intuition behind the analysis is now simple. By \Cref{lem:common-times},
the time span between the endpoint of the left recursive call and the start of
the right recursive call is fixed, independently of the random choices made
inside the recursion. Hence, the transition from the endpoint of $x_\ell$ to the
start of $x_r$ is captured by $P_x$. By \Cref{lem:doubly}, every such
matrix $P_x$ is doubly stochastic. Since products of doubly stochastic matrices
are again doubly stochastic (\Cref{obs:product-ds}), the recursive
identities above allow us to prove inductively that all endpoint-transition
matrices and all leaf-visit matrices are doubly stochastic,
formalized in the upcoming \Cref{lem:recursive-transition-matrices}.

As a consequence, we later obtain that, if the outermost start vertex is
chosen u.a.r. from $S$, then the vertex visited by every fixed leaf is also
distributed u.a.r. over $S$. We also use these matrices to control collisions.
After fixing the complete left subtree at a node $x$, let $u$ be the endpoint
of its left child. Let $(P_x)_u$ denote the row vector of $P_x$ indexed by $u$; 
then $(P_x)_u$ gives, for every vertex $v\in S$, the probability that the right child
starts at $v$. Consequently, for a leaf $y$ in the right subtree, the row
vector $(P_x)_u L_{x_r,y}$ gives the distribution of the vertex visited by
$y$. This distribution is used later to upper bound collisions between the two
subtrees and thereby lower bound the number of new vertices visited. We now
show that all endpoint-transition matrices $E_x$ and all leaf-visit matrices
$L_{x,y}$ are doubly stochastic.

\begin{lemma}\label{lem:recursive-transition-matrices} 
    Let $\cT$ be a recursion tree. For every node $x$ of $\cT$,
    the transition matrix $E_x$ is doubly stochastic, and, for every leaf $y$
    in $\cT_x$, the leaf-visit matrix $L_{x, y}$ is doubly stochastic.
\end{lemma}
\prooflemrecursivetransitionmatrices

Recall that our goal is to construct, using \textsc{PartialExplore}, a walk
$W$ that visits sufficiently many distinct vertices of a fixed set
$D\subseteq S$ of desirable vertices in expectation. We now provide a lower bound on the
expected number of distinct vertices visited in any such set $D$. In the
following, we only analyze vertices visited at leaves of the recursion tree.
This suffices for a lower bound, since vertices visited during the connecting
walks can only
increase the number of distinct vertices visited. We begin by introducing some
additional notation.

For every node $x$ of a recursion tree $\cT$, let $\cL_x$
denote the set of leaves in the subtree $\cT_x$ rooted at $x$. We write
$\cL\coloneq \cL_{x_0}$ for the root $x_0$ of $\cT$. For any node $x$, any set
$\cY\subseteq\cL_x$, and any set $D\subseteq S$, we denote with
$Z_D^{\cY}$ the random variable counting the leaves in
$\cY$ that visit a vertex of $D$, where repeated visits to the same
vertex are counted separately. For the set of all leaves, we write
$Z_D\coloneq Z_D^{\cL}$.

 Let $x_0$ be the root of $\cT$ and let $y$ be an arbitrary leaf.
 Recall that the start vertex of the outermost call is chosen u.a.r.
 from $S$. Therefore, for any fixed vertex $v\in S$, the probability
 that $y$ visits $v$ is $\sum_{p\in S}(1/|S|)(L_{x_0,y})_{p,v}$. By
 \Cref{lem:recursive-transition-matrices}, the leaf-visit matrix $L_{x_0,y}$ is
 doubly stochastic, and thus its column indexed by $v$ sums to $1$. Hence,
 $\Pr[y\text{ visits }v]=1/|S|$ for any choice of $y$ and $v$. Consequently,
 the expected number of leaf visits in $D$, counted with multiplicity, is
 $|\cL|\cdot |D|/|S|$, as formalized in the
 upcoming \Cref{lem:expected-leaf-visits} in a slightly more general way. This
 alone is insufficient to actually visit \textit{different} vertices of $D$:
 the leaves may visit the same vertex of $D$ over and over again. In
 particular, the above statement is true even if, for a u.a.r. start vertex
 $u$, every leaf visits exactly the vertex $u$. We therefore additionally upper
 bound the number of collisions in order to obtain a lower bound
 on the number of distinct vertices of $D$ visited by $W$, as shown later in
 \Cref{lem:coinciding-leaves}.

\begin{lemma}\label{lem:expected-leaf-visits}
    Let $\cY$ be any set of leaves of a recursion tree
    $\cT$. Then, for every $D\subseteq S$, it holds that
    $\bE[Z_D^{\cY}]=|\cY|\cdot |D|/|S|$.
\end{lemma}
\prooflemexpectedleafvisits

Let $D \subseteq S$ be any set of desirable vertices. It remains to give a
lower bound on the expected number of distinct vertices of $D$ visited by an
invocation of \Cref{alg:sample}. Let $\cT$ be a recursion tree
with leaf set $\cL$. We call an unordered pair $\{y,z\}$ of distinct
leaves $y,z\in\cL$ a \emph{collision in $D$} if $y$ and $z$ visit
the same vertex of $D$. We assign every collision to the lowest common
ancestor of its two leaves in $\cT$, and define the random variable
counting the collisions assigned to $x$ with $Q_D^x$. Equivalently,
$Q_D^x$ counts the collisions $\{y,z\}$ with $y\in\cL_{x_\ell}$
and $z\in\cL_{x_r}$. We define $Q_D\coloneq \sum_x Q_D^x$, which is the random
variable counting the total number of collisions. Once we obtain an upper
bound on $\bE[Q_D]$, we can combine it with the lower bound on
$\bE[Z_D]$: for every execution, the number of distinct vertices of $D$
visited at the leaves is at least $Z_D-Q_D$. Since $D$ was chosen arbitrarily
and remains fixed during the analysis for the respective bounds on $Z_D$ and
$Q_D$, the resulting bound holds for every set of desirable vertices. 

\begin{lemma}\label{lem:coinciding-leaves} 
    Let $\cT$ be a recursion tree and let $\cL$ be its set of
    leaves. Then, for any $\delta \in (0, 1)$ and any set $D\subseteq S$, it
    holds that $\bE[Q_D]\leq \delta|\cL|\cdot |D|/|S|$.
\end{lemma}
\prooflemcoincidingleaves

We can now show the following lemma.
\begin{lemma}\label{lem:partial-exploration}
    There is a constant $\gamma>0$ such that \Cref{alg:sample} computes a
    temporal walk $W$ with $\bE[|V(W)\cap D|]\geq \gamma\frac{|D|}{\log |S|}$
    for every set $D\subseteq S$.
\end{lemma}
\prooflempartialexploration

\subsection{Derandomization via Conditional Expectations}\label{ssec:derandom}
We derandomize the randomized choices by the standard method of 
using conditional
expectations; see, e.g., \cite[Chapter~16.1]{AlonS00} for an overview of the
method. Let $\sigma$ be a partial sequence of fixed random choices
of \Cref{alg:sample}. We define
$\Phi_D(\sigma) \coloneq  \bE[Z_D-Q_D\mid \sigma]$.
Recall that $Z_D$ denotes the number of leaf visits in $D$, counted with
multiplicity, and $Q_D$ denotes the number of pairs of distinct leaves that
visit the same vertex of $D$. If $\sigma$ is empty, then
    $\Phi_D(\emptyset) = \bE[Z_D-Q_D]$.
By \Cref{lem:expected-leaf-visits,lem:coinciding-leaves}, $\Phi_D(\emptyset)$ is at
least a constant fraction of $|\cL|\cdot |D|/|S|$, where
$\cL$ is the set of leaves of the recursion tree. Moreover, by
\Cref{lem:number-of-leaves}, we have
$|\cL|=\Theta(K)=\Theta(|S|/\log |S|)$.

The method now fixes the random choices one after another. At every step, we
choose an outcome of the next random choice that does not decrease
$\Phi_D(\sigma)$. Such an outcome exists because the current conditional
expectation is the average of the conditional expectations obtained by fixing
each possible outcome of the next random choice. Hence, after all choices have
been fixed, the resulting deterministic walk satisfies the same lower bound
on $Z_D-Q_D$.

To obtain a polynomial-time derandomization, it remains to prove the following
two claims.

\begin{claim}\label{cl:numchoices}
    \Cref{alg:sample} makes $\mathrm{poly}(n)$ random choices, and
    each such choice has only $\mathrm{poly}(n)$ possible outcomes.
\end{claim}
\proofclnumchoices

\begin{claim}\label{cl:evaluation}
    Let $D \subseteq S$. For every partial fixing $\sigma$,
    the value $\Phi_D(\sigma)$ can be evaluated in $\mathrm{poly}(n)$ time.
\end{claim}

\proofclevaluation

We now simply fix every random choice of the algorithm one by one, at each step
choosing an outcome that does not decrease $\Phi_D(\sigma)$.
\Cref{cl:numchoices} and \Cref{cl:evaluation} give us the runtime guarantee.

\begin{lemma}\label{lem:derandomized}
    \Cref{alg:sample} can be derandomized in polynomial time.
\end{lemma}
\prooflemderandomized

\subsection{Putting it All Together}\label{ssec:together}

A simple iterative application of the previous lemmas gives
\Cref{thm:one-part-exploration}: we repeatedly visit vertices from a set $D$ of
desirable vertices, initially set to $S$, and update $D$ by removing the newly
visited vertices. The procedure was sketched in \Cref{sec:proof-sketch}, see
also \Cref{fig:meta}. The proof follows.
\thmOnePartExploration*
\proofthmonepartexploration

Repeatedly applying \Cref{thm:one-part-exploration} to each fiber
$S\in\cQ$ yields \Cref{thm:exploration}. Using
$\sum_{S\in\cQ}|S|=n$, $|S|\leq n$, and $|\cQ|=q$, the bounds sum to
\[
    \sum_{S\in\cQ} O(q|S|\log^3 |S| + n\log^2 |S|)
    =
    O(qn\log^3 n + qn\log^2 n)
    =
    O(qn\log^3 n).
\]

\thmExploration*

The application to $\trp$ follows as described in \Cref{ssec:trp}.
\thmRendezvous*

\section{Lower Bounds for Temporal Exploration}\label{sec:lower}
The lower bounds of \Cref{cor:lowerbound} follow from two known constructions.
First, consider the lower-bound construction of Erlebach, Hoffmann, and
Kammer~\cite{ErlebachHK21}. They construct a temporal graph
$\deftg=(G_1,\ldots,G_L)$ with vertex set $V=V_1\cup V_2$, where
$|V_1|=|V_2|$, 
such that every snapshot is a star. More precisely, if
$V_1=\{v_1,\ldots,v_{|V_1|}\}$, then the center of the star cycles through the
vertices of $V_1$: at time step $t$, the center is the vertex
$v_{1+((t-1)\bmod |V_1|)}$, while all vertices of $V_2$ are leaves in every
snapshot. This temporal graph requires $\Omega(|V_1|\cdot|V_2|)=\Omega(n^2)$ time steps to be explored.

As shown in~\cite[Lemma~15]{DogeasEKJMW24}, the construction can be
adjusted to obtain, for any $r$ with $1 \le r \le n$, a connected temporal
graph $\tg$ with $r$ automorphism orbits that requires $\Omega(rn)$ time steps for
exploration.
 In the
construction above, all vertices of $V_1$ form singleton orbits, while the
entire set
$V_2$ forms an additional orbit. For $r\le n/2$, we can hence choose
$|V_1|=r-1$; the lower-bound argument is unchanged and
gives $\Omega(rn)$.
For $r>n/2$, we start with the same construction
with
$|V_1|=|V_2|-1$, which
already gives a lower bound of $\Omega(n^2)$. Then we add one final time step,
whose purpose is to split the
orbit $V_2$. If $r=n/2+k$ with $1\le k\le n/2$, this final snapshot consists
of a path containing all vertices of $V_1$ and $k$ vertices of $V_2$, and the
remaining vertices of $V_2$ are attached as pendant vertices to the endpoint of
the path that lies in $V_2$. This splits $V_2$ into $k+1$ orbits, while the
vertices of $V_1$ remain singleton orbits. Hence, the resulting temporal graph
has orbit number $n/2+k=r$. Since the additional snapshot appears at time step
$n^2+1$, the $\Omega(n^2)$ lower bound is preserved, and this is $\Omega(rn)$
for all $r\le n$.

We can express this lower bound in terms of the number $q$ of $2$-WL fibers.
Because every snapshot in the construction above is a tree, already $1$-WL
suffices to compute the automorphism orbits~\cite{KieferSS21}. Hence, the $2$-WL fibers coincide
with the orbits, and the same construction applies with $q$.
This gives a lower bound of $\Omega(qn)$ time steps for every
$1 \leq q \leq n$.

For the lower bound of $\Omega(n \log n)$ for constant $q$, a construction of
Dogeas, Erlebach, Kammer, Meintrup, and Moses Jr.~\cite{DogeasEKJMW24} directly
applies. They showed that there are single-orbit temporal graphs $\deftg$ that
require $\Omega(n\log n)$ time steps for exploration. Their construction can be
extended to any constant number $r$ of orbits by forming $\deftg' = (G_1\square
P_{2r-1},\ldots,G_L\square P_{2r-1})$ where $\square$ denotes the box product
and $P_{2r-1}$ is the path on $2r-1$ vertices. The path factor introduces $r$
different distance classes from the ends of the path. These classes are already
detected by $1$-WL: the endpoints have different degree from the internal path
vertices, and color refinement then propagates this distinction inward along
the path.
    Consequently, the construction above
yields temporal graphs that require $\Omega(n\log
n)$ time steps for exploration whenever $q$ is constant.

} 
\section{Conclusion}\label{sec:conclusion}
We introduced $2$-WL fibers as a structural parameter for temporal exploration
and temporal rendezvous. The main point is that $2$-WL captures enough
information about temporal walks between pairs of vertices to replace the much
stronger use of automorphism orbits. This gives improved bounds and keeps the
resulting algorithms polynomial-time computable.

More generally, our results suggest that the WL algorithm can be used to study the
expressiveness of temporal graphs with respect to reachability. Previous work
compares temporal graph settings by the reachability graphs they can express,
yielding a hierarchy of temporal reachability
models~\cite{CasteigtsCS24,Doring25}. In this context, it would be interesting
to understand which parts of the temporal-reachability hierarchy can be explained by WL equivalence.
A particularly natural part of the reachability hierarchy is that of
\emph{proper} temporal graphs, where no two incident edges appear at the same
time step. Since temporal reachability is a relation between ordered pairs of
vertices, $2$-WL is the lowest-dimensional version of the WL algorithm that directly reflects such a pairwise
structure. This suggests the following direction. We expect that proper
temporal graphs are the setting in which reachability is exactly captured by
$2$-WL, in the sense that pairs indistinguishable by $2$-WL have the same
temporal-reachability behavior. Understanding whether this is true, and how it
extends to other types of temporal graphs within this hierarchy would clarify
the potential of the WL algorithm as a measure of temporal-reachability expressiveness.

Another direction is to study agent problems on temporal graphs through the
lens of WL. Our rendezvous result shows that the structure exposed by $2$-WL
fibers is sufficient for two agents without a common labeling to meet quickly.
This suggests that other temporal agent problems may also admit structural
descriptions in terms of WL fibers or related equivalence notions. This is in
line with the role of WL in game-based characterizations, such as the bijective
pebble game~\cite{GroheM12}, and its connections to cops-and-robber games in
lower bound constructions~\cite{GroheLNS25}. More broadly, $k$-WL and the
$k$-WL fibers should be studied as a structural parameter for temporal-walk
problems such as reachability, connectivity, temporal spanning structures, and
agent-based games, such as temporal cops and robbers
variants~\cite{MorawietzRW20}.

Finally, our approaches for \texp{} and \trp{} rely on
information about pairs of vertices and in particular on the counts of temporal walks
between such pairs. This is why $2$-WL fits the scenario naturally. It would
nevertheless be interesting to understand where exactly $1$-WL (or a slightly enhanced version of it) has its limitations for these problems, given that it does compute degrees between vertex fibers. We anticipate that such an approach would require a substantially different analysis than ours.

\section*{Acknowledgements}
\new{This work has benefited from Dagstuhl Seminar 24441,
``Machine Learning Augmented Algorithms for Combinatorial Optimization Problems'', where the research was initiated.

ChatGPT (GPT-6) was used for limited language editing and to assist with an extended search for relevant literature. All scientific ideas, claims, and mathematical proofs are the original work of the authors, who take full responsibility for the content.}

\ifarxiv
    \bibliographystyle{plainnat}
\fi
\bibliography{main}
\end{document}